\documentclass[aps,pra,floatfix,superscriptaddress,preprint,showpacs]{revtex4}

\pdfoutput=1

\usepackage{graphicx, amsmath, amssymb} 

\newcommand{\al}{\alpha}     
      
\newcommand{\ga}{\gamma}         
\newcommand{\de}{\delta}         
\newcommand{\ep}{\epsilon}

\newcommand{\et}{\eta}       
\newcommand{\thet}{\theta}

\newcommand{\ka}{\kappa}     \newcommand{\iKa}{K}

\newcommand{\rh}{\rho}

\newcommand{\ph}{\phi}             

\newcommand{\ch}{\chi}       \newcommand{\iCh}{X}
\newcommand{\ps}{\psi}             
\newcommand{\om}{\omega}         
\newcommand{\es}{\;\;\;\:\;}
\newcommand{\beq}{\begin{equation}}
\newcommand{\eeq}{\end{equation}}
\newcommand{\beqs}{\begin{equation*}}
\newcommand{\eeqs}{\end{equation*}}
\newcommand{\bea}{\begin{equation}\begin{aligned}}
\newcommand{\eea}{\end{aligned}\end{equation}}
\newcommand{\beaa}{\begin{equation}\begin{alignedat}{2}}
\newcommand{\eeaa}{\end{alignedat}\end{equation}}
\newcommand{\beaaa}{\begin{equation}\begin{alignedat}{3}}
\newcommand{\eeaaa}{\end{alignedat}\end{equation}}
\newcommand{\bean}[1]{\begin{equation}\begin{alignedat}{#1}}
\newcommand{\eean}{\end{alignedat}\end{equation}}
\newcommand{\beg}{\begin{equation}\begin{gathered}}
\newcommand{\eeg}{\end{gathered}\end{equation}}
\newcommand{\bem}{\begin{multline}}
\newcommand{\eem}{\end{multline}}

\newcommand{\bm}[1]{\begin{bmatrix}#1\end{bmatrix}}
\newcommand{\bmb}{\begin{bmatrix}}
\newcommand{\emb}{\end{bmatrix}}
\newcommand{\p}[1]{\left(#1\right)}
\newcommand{\pb}[1]{\left[#1\right]}

\newcommand{\pv}[1]{\left|#1\right|}
\newcommand{\pV}[1]{\left\Vert#1\right\Vert}
\newcommand{\pS}[1]{\left(#1\right)_S}

\newcommand{\<}[1]{\p{#1}}
\newcommand{\C}[2]{\pb{#1,#2}}

\newcommand{\E}[1]{\left<#1\right>}

\newcommand{\BOK}[3]{\left<#1\middle|#2\middle|#3\right>}

\newcommand{\te}[1]{{#1}}
\newcommand{\ma}[1]{\mathbf{#1}}
\newcommand{\gma}[1]{\boldsymbol{#1}}

\newcommand{\as}[1]{\mathbb{#1}}
\newcommand{\un}[1]{\mathrm{#1}}

\newcommand{\txt}[1]{\textrm{#1}}
\newcommand{\ham}{\mathcal{H}}

\newcommand{\ee}[1][]{{\mathrm{e}^{#1}}}
\newcommand{\ii}{\mathrm{i}}

\newcommand{\hb}{\hbar}

\newcommand{\ct}{^{\dagger}}
\newcommand{\pr}{^{\prime}}

\newcommand{\abs}[1]{\pv{#1}}
\newcommand{\norm}[2][]{\pV{#2}_{#1}}
\DeclareMathOperator{\sgn}{sgn}

\newcommand{\dd}{\mathrm{d}}

\newcommand{\dz}{\dd z}

\newcommand{\D}[3][]{\frac{\dd^{#1}#2}{\dd {#3}^{#1}}}
\newcommand{\Dt}[2][]{\D[#1]{#2}{t}}

\DeclareMathOperator{\trace}{Tr}
\newcommand{\tr}[2][]{\trace_{#1}\left\{#2\right\}}
\newcommand{\dm}{\te{\rh}}
\newcommand{\aop}[1][]{\te{a_{#1}}}
\newcommand{\ad}[1][]{{\te{a_{#1}}\ct}}
\newcommand{\ada}[1][]{\ad[#1]\aop[#1]}

\newcommand{\bop}[1][]{\te{b_{#1}}}
\newcommand{\bd}[1][]{{\te{b_{#1}}\ct}}
\newcommand{\bdb}[1][]{\bd[#1]\bop[#1]}

\newcommand{\f}[2][1]{\frac{#1}{#2}}

\newcommand{\eit}[1]{\ee[\ii {#1}t]}
\newcommand{\emit}[1]{\ee[-\ii {#1}t]}

\begin{document}

\title{Quantum entanglement between a nonlinear nanomechanical resonator and a microwave field.}

\author{Charles P Meaney}
\affiliation{Centre for Quantum Computer Technology, School of Mathematical and Physical Sciences, The University of Queensland, St Lucia, QLD 4072, Australia}
\author{Ross H McKenzie}
\affiliation{ Department of Physics, School of Mathematical and Physical Sciences, The University of Queensland, St Lucia, QLD 4072, Australia}
\author{G J Milburn}
\affiliation{Centre for Quantum Computer Technology, School of Mathematical and Physical Sciences, The University of Queensland, St Lucia, QLD 4072, Australia}

\begin{abstract}
 We consider a theoretical model for a nonlinear nanomechanical resonator coupled to a superconducting microwave resonator. The nanomechanical resonator is driven parametrically at twice its resonance frequency, while the superconducting microwave resonator is driven with two tones that differ in frequency by an amount equal to the parametric driving frequency. We show that the semi-classical approximation of this system has an interesting fixed point bifurcation structure. In the semi-classical dynamics a transition from stable fixed points to limit cycles is observed as one moves from positive to negative detuning. We show that signatures of this bifurcation structure are also present in the full dissipative quantum system and further show that it leads to mixed state entanglement between the nanomechanical resonator and the microwave cavity in the dissipative quantum system that is a maximum close to the semi-classical bifurcation. Quantum signatures of the semi-classical limit-cycles are presented.
\end{abstract}

\maketitle

\section{Introduction.}

\label{s:i}
The steady states of driven, dissipative quantum systems can manifest entanglement between the component  physical systems\cite{Schneider,Kraus,Verstraete, Esslinger,Cho,Orth}. In such cases quantum correlations survive despite dissipation, although the resulting states are not necessarily pure.  Steady state entanglement, in contrast to entanglement through unitary dynamics, does not depend on the initial state. For these reasons, the prospect of engineering driven dissipative systems so that their steady states exhibit  some desirable entanglement is an intriguing prospect\cite{Diehl}. 

In recent years, another aspect of engineered quantum systems has arisen in the context of circuit quantum electrodynamics (circuit QED)\cite{devoret:2007}, nano-mechanics\cite{schwab:2005} and optomechanics\cite{optomechanics-review-Science}. A characteristic feature of the description of these systems is an effective quantisation in which quantum theory is used to describe collective macroscopic degrees of freedom rather than atomic degrees of freedom. Typical examples include the voltage and current in an equivalent circuit description of a a circuit containing superconducting junctions or a bulk flexural mode of a nanomechanical resonator. This approach works because at low temperatures these collective degrees of freedom largely factor out of the microscopic degrees of freedom which remain only in so far as a source of dissipation and noise.  

Of particular interest in this paper  is an experimental context involving nano-scale resonators coupled to superconducting coplanar waveguides, these devices are made from aluminium on a silicon substrate, and are placed in a dilution refrigerator. There is ongoing effort to cool these nanomechanical systems to close to the ground state of a collective bulk flexural mode where quantum mechanical phenomena such as entanglement become manifest. Rapid experimental progress means that such a quantum regime is now accessible\cite{Cleland2010}. In this paper we propose a particular kind of engineered dissipative quantum system, a nonlinear nanomechanical resonator coupled to a superconducting microwave resonator, that exhibits an entangled dissipative steady state. The correlations implicit in the entangled state are enforced by the correlations between fixed points of the corresponding semiclassical dynamical system. 

A nanomechanical resonator can form one plate of a capacitor coupling a coherent driving field to the cavity field\cite{teufel:2008,woolley:2008}. The nanomechanical element can also exhibit a significant Duffing nonlinearity \cite{almog:2007}. Such properties allow the investigation of a nonlinear nanomechanical system. The classical model possesses distinct nonlinear phenomena in its steady state, and we seek a signature of these in the quantum system.

The paper is structured as follows. In section \ref{s:model} we introduce the nonlinear nanomechanical system considered in this paper. We give a description in terms of an effective Hamiltonian and a Markov master equation describing dissipation of both microwave and mechanical modes.  In section \ref{s:sc} we present a detailed analysis of the semiclassical dissipative dynamics of this system in terms of the fixed point stability.  In section \ref{s:q} we consider the quantum version of the model. We numerically determine the steady state of the system in the number basis for both resonators, suitably truncated. The steady state entanglement between the mechanics and the field can then be calculated in terms of the log negativity.  From the steady state density matrix in the number basis, we construct the marginal Wigner functions for mechanical and cavity degrees of freedom. We  show that, as the control parameters are varied through the values at which the semiclassical model shows bifurcations, the Wigner functions become double peaked with dominant support on the semiclassical fixed points. This enables us to elucidate the nature of the entanglement in the quantum steady state.  Finally in section \ref{s:c} we summarise our results and suggest new directions for further work.

\section{The dissipative Cassinian oscillator model}

\label{s:model}

We consider a nonlinear nanomechanical resonator coupled to a superconducting microwave resonator. The nanomechanical nonlinearity is of the well-established Duffing (quartic) type \cite{kozinsky:2007}. We start by defining the field mode annihilation operator $\te{a}$ of the superconducting microwave resonator at frequency $\om_c$. We also define two field quadrature operators $\te{x_a}$ and $\te{y_a}$ as
\bea
 \te{x_a} & = \f{2}\<{\aop+\ad},     & \aop & = \te{x_a}+\ii\te{y_a}, \\
 \te{y_a} & = -\ii\f{2}\<{\aop-\ad}, & \ad  & = \te{x_a}-\ii\te{y_a} \,.
\eea
Similarly, we define the mechanical mode annihilation operator $\te{b}$ at frequency $\om_m$. The oscillator mass is defined here to be $m$, and we also define the position and momentum operators $\te{q}$ and $\te{p}$ as
\bea
 \te{q} & = \sqrt{\f[\hb]{2m\om_m}}\<{\bop+\bd},      & \bop & = \f{\sqrt{2\hb}}\<{\sqrt{m\om_m}\te{q}+\ii\f{\sqrt{m\om_m}}\te{p}}, \\
 \te{p} & = -\ii\sqrt{\f[\hb m\om_m]{2}}\<{\bop-\bd}, & \bd  & = \f{\sqrt{2\hb}}\<{\sqrt{m\om_m}\te{q}-\ii\f{\sqrt{m\om_m}}\te{p}} \,;
\eea
as well as two normalised operators $\te{x_b}$ and $\te{y_b}$ as
\bea
 \te{x_b} & = \f{2}\<{\bop+\bd},     & \bop & = \te{x_b}+\ii\te{y_b}, \\
 \te{y_b} & = -\ii\f{2}\<{\bop-\bd}, & \bd  & = \te{x_b}-\ii\te{y_b} \,.
\eea
The relevant commutation relations for the microwave and mechanical resonators are
\bea
 \C{\aop}{\ad} & = \te{I}, & \C{\te{x_a}}{\te{y_a}} & = \ii\f{2}\te{I}, \\
 \C{\bop}{\bd} & = \te{I}, & \C{\te{x_b}}{\te{y_b}} & = \ii\f{2}\te{I}, & \C{\te{q}}{\te{p}} & = \ii\hb\te{I} \,,
\eea
where $\te{I}$ is the identity.

The capacitive coupling between the nanomechanical resonator and the cavity field is such as to shift the cavity frequency by an amount proportional to the displacement of the nanomechanical resonator. As the bare cavity frequency is very high (GHz) compared to the observational time scales of interest we can approximate the interaction so that it is proportional to the number operator for the cavity field times the mechanical displacement. The mechanical resonator thus acts as a phase modulator for the cavity field. Such a coupling can also be achieved in an opto-mechanical context where radiation pressure moves one mirror of an optical cavity; see \cite{pace:1993} for a derivation. 

In our model, the microwave cavity is coherently driven with two tomes, as in \cite{clerk:2008}; and the mechanical oscillator is driven parametrically. Such a parametrically driven oscillator can achieve squeezing which is maximal at the threshold for parametric oscillation \cite{walls:2007}. Finally,  we include in our model of the nanomechanical  an anharmonic component. The dominant nonlinearity is quartic (in the Hamiltonian, cubic in the force) in the displacement of the nanomechanical resonator,  and is known as the ``Duffing'' nonlinearity; a derivation for nanomechanical resonators is found in \cite{lifshitz:2008}.

\begin{figure}[!htbp]\begin{center}
\includegraphics[scale=0.75]{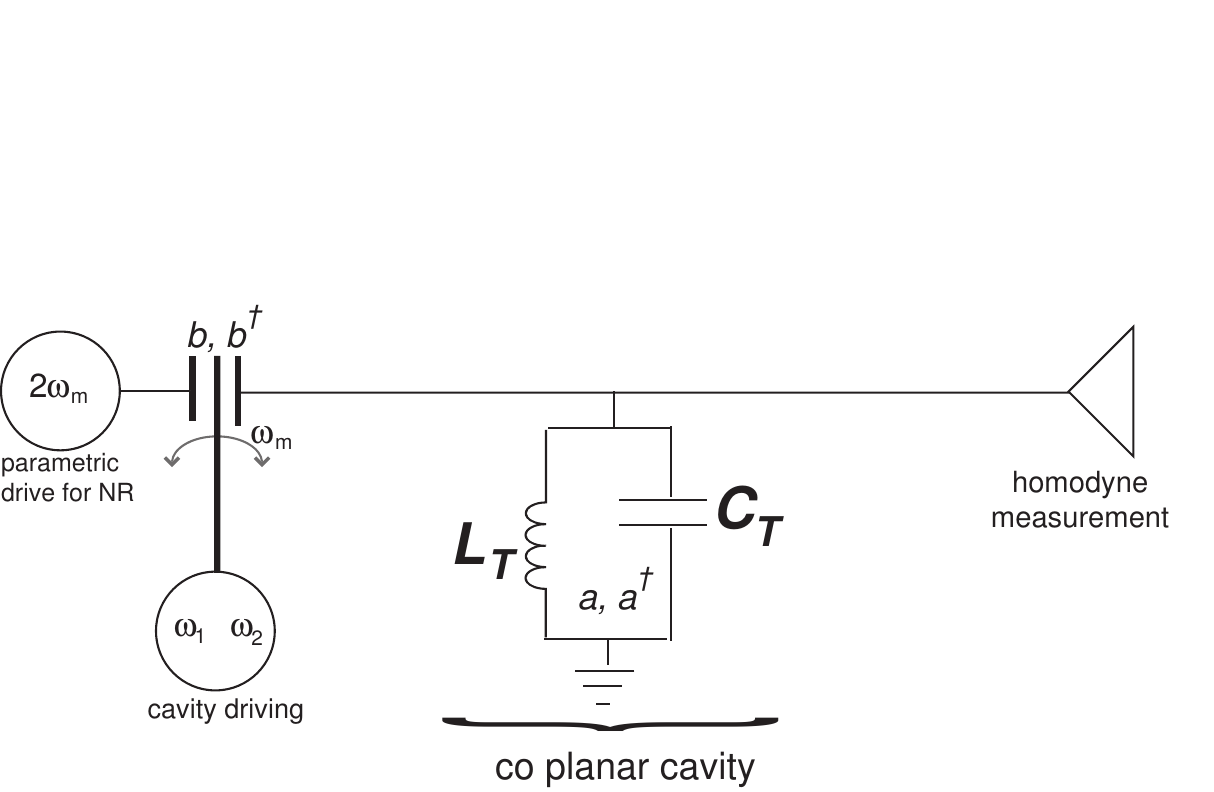}
\caption{A schematic of the system under consideration. A nonlinear nanomechanical resonator (NR) of frequency $\om_m$ is coupled to a superconducting microwave resonator of frequency $\om_c$. The nanomechanical oscillator is driven parametrically at frequency $2\om_m$; and the microwave cavity is driven at two tones $\om_1$ and $\om_2$. The annihilation operators for the microwave and nanomechanical modes are $\aop$ and $\bop$, respectively. In terms of these operators, the coupling takes the form $\ad\aop\<{\bop+\bd}$. The superconducting microwave resonator is modelled as a lumped element LC circuit, where $L_T$ and $C_T$ are the inductance and capacitance of the tank circuit respectively.}
\label{f:system_setup}
\end{center}\end{figure}

\begin{figure}[!htbp]\begin{center}
\includegraphics[scale=0.75]{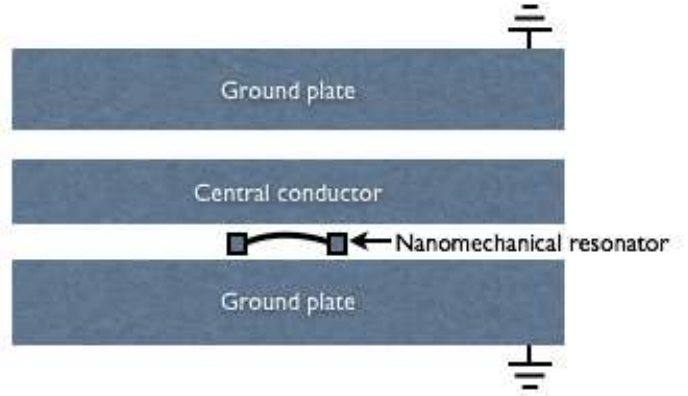}
\caption{A physical schematic of the system under consideration. A nonlinear nanomechanical oscillator of frequency $\om_m$ is capacitively coupled to a superconducting microwave resonator of frequency $\om_c$. A theoretical schematic of the system is shown in Figure \ref{f:system_setup}.}
\label{f:phys_setup}
\end{center}\end{figure}

 We thus have the Schr\"odinger picture Hamiltonian describing the coupled system
\begin{multline}
 \ham = \hb\om_c\ada + \hb\sum_{i=1}^2\<{\ep_i^*\aop\eit{\om_i}+\ep_i\ad\emit{\om_i}} + \f[m\om_m^2]{2}\te{q}^2 + \f{2m}\te{p}^2 + \iCh\te{q}^4  \\
        + \iKa\cos\<{2\om_mt}\te{q}^2 + \hb G_0\ada\te{q} \,,
\end{multline}
where: $\om_c$ is the frequency of the microwave cavity; $\ep_1$ and $\ep_2$ are the amplitudes of the two linear drives on the microwave cavity; $m$ and $\om_m$ are the mass and frequency of the mechanical oscillator respectively; $\iCh$ is the amplitude of the Duffing nonlinearity of the mechanical resonator; $\iKa$ is the amplitude of the parametric pumping of the nonlinear mechanical resonator; and $G_0$ is the amplitude of the coupling between the photon number of the cavity and the position of the mechanical resonator. A derivation of the form of the capacitive coupling between microwave resonator and mechanical oscillator as $\ada\te{q}$ is given in \cite{woolley:2008}; a typical achievable value for the coupling strength is $G_0=2\pi\times1.16\un{kHz nm^{-1}}$ \cite{lehnert:2008}. An equivalent circuit schematic of the system is given in Figure \ref{f:system_setup}. A physical schematic of the microwave-nanomechanical coupling is given in Figure \ref{f:phys_setup}. The reader is referred to \cite{teufel:2008b} for a scanning electron microscope (SEM) image of such a device. Nanomechanical Duffing oscillators have been discussed in \cite{kozinsky:2007,babourina-brooks:2008}; parametric excitation in the nano-mechanical context has been discussed in \cite{woolley:2008,hertzberg:2010,rocheleau:2010}.

We now move to an interaction picture in a rotating frame with respect to the average of the linear drive frequencies in the microwave resonator space, and the mechanical resonator frequency in its space. We set $\f[\om_2-\om_1]{2}=\om_m$ and make the rotating wave approximation by ignoring terms with frequency $2\om_m$ or above. Next, we linearise about the steady state, using the same ansatz as Woolley et al \cite{woolley:2008}, and set $\ep_1=\ep\ee[-\ii\ps]$ and $\ep_2=-\ep\ee[\ii\ps]=-\ep_1^*$ where $\ep\in\as{R}$. Finally, we choose the phase $\ps=\f[\pi]{2}$ to obtain the effective Hamiltonian
\beq
 \tilde\ham_E = \hb\om\ada + \hb\f[\ch]{2}\pS{\bop^2\bd^2} + \hb\f[\ka]{2}\<{\bop^2+\bd^2} - \ii\hb\f[g]{2}\<{\aop+\ad}\<{\bop-\bd} \,,
\eeq
where the notation $\pS{\cdots}$ is used for a symmetrised product, such that
\beq
 \pS{\bop^2\bd^2} = \f{6}\<{\bop^2\bd^2 + \bop\bd\bop\bd + \bop\bd^2\bop + \bd\bop^2\bd + \bd\bop\bd\bop + \bd^2\bop^2} \,,
\eeq
the cavity detuning $\om$ is
\beq
 \om = \om_c-\f[\om_1+\om_2]{2} \,,
\eeq
and $\ch$, $\ka$, and $g$ are the rescaled mechanical Duffing nonlinearity, parametric pumping amplitude, and cavity-mechanical coupling respectively, where
\bea
 \ch & = \f[12\hb]{\<{2m\om_m}^2}\iCh, & \ka & = \f{2m\om_m}\iKa, & g & = -\f[2\ep]{\om_m}\sqrt{\f[\hb]{2m\om_m}}\;G_0 \,.
\eea
Classical trajectories from the nanomechanical portion of this Hamiltonian are the ovals of Cassini, and the mechanical system is hence sometimes described as the ``Cassinian'' oscillator; the quantum version has been previously studied by Wielinga et al \cite{wielinga:1993}. A similar model has been extensively discussed by Dykmann and his collaborators\cite{Dykmann} and also by Peano and Thowart\cite{PW}. A typical value for the Duffing nonlinearity is $\ch\sim6.8\times10^{-4}\un{s^{-1}}$ \cite{kozinsky:2007}.

For a realistic description we adopt a dissipative model. We model both the microwave cavity resonator and the mechanical resonator as being damped in zero temperature heat baths. A zero temperature heat bath for the cavity is certainly justified as the typical microwave cavity is at mK temperature and thus very close to zero \cite{wallraff:2004}. The zero temperature heat bath for the nanomechanics is not as good an approximation. However, the mean thermal occupation of the bath $\bar{n}\ne0$ does not enter the semi-classical equations, and thus the semi-classical bifurcation structure studied in section \ref{s:sc} will be the correct one. Yet $\bar{n}\ne0$ does affect the quantum steady state and the quantum entanglement calculation. For this paper however, we do not study the temperature dependence of the steady state entanglement. We thus consider the zero temperature heat bath for the nanomechanics to be satisfactory for the purpose of showing the classical-quantum correspondence of sections \ref{s:sc} and \ref{s:q}. The amplitude decay for the microwave cavity is $\mu$, and for the nanomechanical resonator is $\ga$. We then describe the dissipative dynamics with the master equation (with weak damping and the rotating wave approximation for the system-environment couplings)
\beq
 \Dt{\dm} = -\f[\ii]{\hb}\C{\tilde\ham_E}{\dm} + \mu\<{2\aop\dm\ad-\ada\dm-\dm\ada} + \ga\<{2\bop\dm\bd-\bdb\dm-\dm\bdb} \,,
 \label{e:me}
\eeq
where $\dm$ is the density matrix of the coupled system.

\section{Semi-classical fixed point structure}

\label{s:sc}

\subsection{Reparameterisation}

For the model defined by the master equation \eqref{e:me}, we will now find semi-classical steady states. We assume that all coupling parameters are non-zero ($\om,\ch,\ka,g\ne0$), and that the two dissipation parameters are also positive ($\mu,\ga>0$).

First, we define four convenient dimensionless parameters: a renormalised inverse square microwave-nanomechanical coupling parameter $\xi$; a renormalised parametric drive parameter $\ka\pr$; a renormalised microwave cavity dissipation parameter $\mu\pr$; and a renormalised nanomechanical dissipation parameter $\ga\pr$. These are defined by
\bea
 \xi    & = \f[2\<{\om^2+\mu^2}]{g^2} \,, \\
 \ka\pr & = \f[\ka]{\om} \,, \\
 \mu\pr & = \f[\mu]{\om} \,, \\
 \ga\pr & = \f[\ga]{\om} \,.
\eea
In terms of our previous assumptions, all of these new parameters are non-zero ($\ka\pr,\mu\pr,\ga\pr\ne0$), and the renormalised coupling is also positive ($\xi>0$). Additionally, we note that since the original dissipation parameters $\mu$ and $\ga$ are always positive, the sign of the corresponding renormalised parameters is equal to the sign of the detuning $\om$; specifically $\sgn\<{\om}=\sgn\<{\mu\pr}=\sgn\<{\ga\pr}$. In what follows we will write $\sgn\<{\om}$ whenever this value is required to remind us of its physical origin. We also define one additional parameter $\ph$, which we will see determines the locations of the semi-classical fixed point bifurcations,
\beq
 \ph = \<{1+\xi\ka\pr}^2 - \<{\xi\ga\pr}^2 \,. \label{e:phdef}
\eeq

In the consideration of the semi-classical fixed point structure, it turns out to be most useful to consider the re-scaling of the mode operators
\bea
 \aop & \to \f[\sqrt{2}\,\et]{\sqrt{\xi}}\bar{\aop} \,, \\
 \bop & \to \et\bar{\bop} \,,
\eea
where
\beq
 \et = \sqrt{\f[\ka]{\ch}\f{\xi\ka\pr}\<{1+\sgn\<{\om}\sqrt{\ph}}} \,,
\eeq
and thus the commutation relations become
\bea
 \C{\bar\aop}{\bar\aop\ct} & = \f[\om^2+\mu^2]{\et^2g^2}\te{I}, \\
 \C{\bar\bop}{\bar\bop\ct} & = \f{\et^2}\te{I} \,.
\eea
This re-parametrisation is a real scaling if $\et>0$, which is true if both
\bea
 \ph & > 0 & \txt{and} && \sgn\<{\ch}\<{\sgn\<{\om}+\sqrt{\ph}} & > 0 \,.
 \label{e:rc}
\eea
Performing this re-parametrisation, the master equation \eqref{e:me} becomes
\beq
 \Dt{\dm} = \f[2\et^2\om]{\xi}\<{-\ii\C{\bar\ham}{\dm} + \mu\pr\<{2\bar\aop\dm\bar\aop\ct-\bar\aop\ct\bar\aop\dm-\dm\bar\aop\ct\bar\aop} + \f[\xi\ga\pr]{2}\<{2\bar\bop\dm\bar\bop\ct-\bar\bop\ct\bar\bop\dm-\dm\bar\bop\ct\bar\bop}} \,,
 \label{e:mep}
\eeq
where
\begin{multline}
 \bar\ham = \bar\aop\ct\bar\aop + \f[1 + \sgn\<{\om}\sqrt{\ph}]{4}\pS{{\bar\bop}^2{\bar\bop}^{\dagger2}} + \f[\xi\ka\pr]{4}\<{{\bar\bop}^2+{\bar\bop}^{\dagger2}} \\ {}- \ii\f[\sgn\<{\om}\sqrt{1+{\mu\pr}^2}]{2}\<{\bar\aop+\bar\aop\ct}\<{\bar\bop-\bar\bop\ct} \,.
\end{multline}

We now define new quadrature operators for the re-parametrised modes as
\bea
 \bar{\te{x_a}} & = \f{2}\<{\bar\aop+\bar\aop\ct} \,, \\
 \bar{\te{y_a}} & = -\ii\f{2}\<{\bar\aop-\bar\aop\ct} \,, \\
 \bar{\te{x_b}} & = \f{2}\<{\bar\bop+\bar\bop\ct} \,, \\
 \bar{\te{y_b}} & = -\ii\f{2}\<{\bar\bop-\bar\bop\ct} \,.
\eea
The quantum equations of motion for both quadratures of both oscillators are then obtained from the re-parametrised master equation model \eqref{e:mep}, and are found to be
\bea
 \Dt{\E{\bar{\te{x_a}}}} & =   \f[\xi]{2\et^2}\pb{-\mu\pr\E{\bar{\te{x_a}}} + \E{\bar{\te{y_a}}}} \,, \\
 \Dt{\E{\bar{\te{y_a}}}} & =   \f[\xi]{2\et^2}\pb{-\E{\bar{\te{x_a}}} - \mu\pr\E{\bar{\te{y_a}}} + \sgn\<{\om}\sqrt{1+{\mu\pr}^2}\E{\bar{\te{y_b}}}} \,, \\
 \Dt{\E{\bar{\te{x_b}}}} & =   \f{2\et^2}\pb{\sgn\<{\om}\sqrt{1+{\mu\pr}^2}\E{\bar{\te{x_a}}} - \xi\ga\pr\E{\bar{\te{x_b}}} - \xi\ka\pr\E{\bar{\te{y_b}}} \right. \\
                   & \es \left. {}+ \<{1 + \sgn\<{\om}\sqrt{\ph}}\<{\E{\pS{\bar{\te{x_b}}^2\bar{\te{y_b}}}} + \E{\bar{\te{y_b}}^3}}} \,, \\
 \Dt{\E{\bar{\te{y_b}}}} & =   \f{2\et^2}\pb{-\xi\ga\pr\E{\bar{\te{x_b}}} - \xi\ga\pr\E{\bar{\te{y_b}}} - \<{1 + \sgn\<{\om}\sqrt{\ph}}\<{\E{\bar{\te{x_b}}^3} + \E{\pS{\bar{\te{x_b}}\bar{\te{y_b}}^2}}}} \,.
\eea
These equations of motion for the expectations are seen to couple to the equations of motion of third order moments, which in turn couple to infinite orders. The only case where this does not happen would be if $1 + \sgn\<{\om}\sqrt{\ph} = 0$, and this case is prevented by our conditions for a valid real re-parametrisation \eqref{e:rc}.

\subsection{Semi-classical model}

To proceed, we define the semi-classical equations by factorising the third order moments, and thus obtain a closed system of equations. Specifically, we make the four assumptions
\bea
 \E{\bar{\te{x_b}}^3}                      & = \E{\bar{\te{x_b}}}^3 \,, \\
 \E{\pS{\bar{\te{x_b}}^2\bar{\te{y_b}}}}   & = \E{\bar{\te{x_b}}}^2\E{\bar{\te{y_b}}} \,, \\
 \E{\pS{\bar{\te{x_b}}^3\bar{\te{y_b}}^2}} & = \E{\bar{\te{x_b}}}\E{\bar{\te{y_b}}}^2 \,, \\
 \E{\bar{\te{y_b}}^3}                      & = \E{\bar{\te{y_b}}}^3 \,.
\eea
This is equivalent to saying that the covariances are much less than the corresponding product of the means in each case. After factorising the third order moments, four semi-classical variables are then defined by
\bea
 \E{\bar{\te{x_a}}} & \mapsto x_a \,, \\
 \E{\bar{\te{y_a}}} & \mapsto y_a \,, \\
 \E{\bar{\te{x_b}}} & \mapsto x_b \,, \\
 \E{\bar{\te{y_b}}} & \mapsto y_b \,.
\eea
These replacements generate the semi-classical equations of motion
\bea
 \Dt{x_a} & =   \f[\xi]{2\et^2}\pb{-\mu\pr x_a + y_a} \,, \\
 \Dt{y_a} & =   \f[\xi]{2\et^2}\pb{-x_a - \mu\pr y_a + \sgn\<{\om}\sqrt{1+{\mu\pr}^2}y_b} \,, \\
 \Dt{x_b} & =   \f{2\et^2}\pb{\sgn\<{\om}\sqrt{1+{\mu\pr}^2}x_a - \xi\ga\pr x_b - \xi\ka\pr y_b + \<{1 + \sgn\<{\om}\sqrt{\ph}}\<{x_b^2y_b + y_b^3}} \,, \\
 \Dt{y_b} & =   \f{2\et^2}\pb{-\xi\ka\pr x_b - \xi\ga\pr y_b - \<{1 + \sgn\<{\om}\sqrt{\ph}}\<{x_b^3 + x_by_b^2}} \,.
 \label{e:sceom}
\eea
These semi-classical equations of motion \eqref{e:sceom} have fixed points ($\Dt{x_a} = \Dt{y_a} = \Dt{x_b} = \Dt{y_b} = 0$) which are the steady states of the semi-classical dynamical system if they are stable. Thus, we must also consider the stability of the fixed points. The defined semi-classical variables can be considered as a vector $\ma{x}$, such that about a fixed point $\ma{x_0}$ we have
\beq
 \gma{\de}\ma{x} = \ma{x}-\ma{x^0} = \pb{x_a-x_{a_0},y_a-y_{a_0},x_b-x_{b_0},y_b-y_{b_0}}^T \,,
\eeq
\beq
 \Dt{}\gma{\de}\ma{x} = \ma{M}\gma{\de}\ma{x} \,,
\eeq
where the Jacobian matrix $\ma{M}$ is
\beq
 \ma{M} = \f{\et^2}\bm{
   -\f[\xi]{2}\mu\pr                   & \f[\xi]{2}        & 0                                                                      & 0 \\
   -\f[\xi]{2}                         & -\f[\xi]{2}\mu\pr & 0                                                                      & \f[\xi]{2}\sgn\<{\om}\sqrt{1+{\mu\pr}^2} \\
   \f{2}\sgn\<{\om}\sqrt{1+{\mu\pr}^2} & 0                 & -\f[\xi\ga\pr]{2} + \<{1 + \sgn\<{\om}\sqrt{\ph}}x_by_b                & -\f[\xi\ka\pr]{2} + \f[1 + \sgn\<{\om}\sqrt{\ph}]{2}\<{x_b^2 + 3y_b^2} \\
   0                                   & 0                 & -\f[\xi\ka\pr]{2} - \f[1 + \sgn\<{\om}\sqrt{\ph}]{2}\<{3x_b^2 + y_b^2} & -\f[\xi\ga\pr]{2} - \<{1 + \sgn\<{\om}\sqrt{\ph}}x_by_b} \,.
\eeq
Stability of the fixed point requires all the eigenvalues of the Jacobian to have a real part less than or equal to zero \cite{hilborn:1994}. A real part of exactly zero indicates marginal stability in that parameter direction, where the fixed point is neither attractive nor repulsive. Real parts strictly less than zero are attracting fixed points which draw in nearby regions in phase space. In general, stability may depend on more coupling parameter combinations than those which define the fixed points. This is indeed the case here, and will be discussed explicitly in the next section.


\subsection{Semi-classical fixed points}

There are three different classes of fixed points of the semi-classical equations of motion \eqref{e:sceom}. The fixed points of the semi-classical equations of motion for the microwave resonator field, the first two expressions of \eqref{e:sceom}, give the relationship between the nanomechanical steady phase space and the microwave resonator field. Explicitly, for all three classes of semi-classical fixed points we have,
\bea
 x_{a_0} & = -\sgn\<{\om}\cos\<{\arctan\mu\pr}y_{b_0} \,, \\
 y_{a_0} & = -\sgn\<{\om}\sin\<{\arctan\mu\pr}y_{b_0} \,.
 \label{e:cncorrelation}
\eea
These relations force a correlation between the mechanical and electromagnetic resonator at the fixed points. This will be significant when we consider entanglement in the quantum steady state in section \ref{s:q}.

Next, to find the scaled momentum of the nanomechanical steady state $y_{b_0}$, and to express the nanomechanical components of the steady state in general, it is better to use polar coordinates. We define polar coordinates for the nanomechanical phase space $r_{b_0}$ and $\thet_{b_0}$ in the usual way
\bea
 x_{b_0} & = r_{b_0}\cos\thet_{b_0} \,, \\
 y_{b_0} & = r_{b_0}\sin\thet_{b_0} \,.
\eea
In polar coordinates, the nanomechanical components of the three classes of semi-classical fixed points are
\bea
 \<{r_{b_0}^2,\tan\thet_{b_0}} & =   \<{0,\cdot}, \\
                               & \es \<{\f[1+\sqrt{\ph}]{1+\sgn\<{\om}\sqrt{\ph}}, \f[-1-\xi\ka\pr-\sqrt{\ph}]{\xi\ga\pr}}, \\  
                               & \es \<{\f[1-\sqrt{\ph}]{1+\sgn\<{\om}\sqrt{\ph}}, \f[-1-\xi\ka\pr+\sqrt{\ph}]{\xi\ga\pr}} \,. 
 \label{e:scfp}
\eea

The first class of fixed points is the trivial solution at the origin where both quadratures of both the microwave and nanomechanical resonators are zero. The fixed point at the origin is sometimes, though not always, stable as will be discussed further below. The second and third classes of fixed points are both pairs of fixed points at anti-podal positions in the nanomechanical phase space. In both cases, the microwave resonator phase space components are also a pair of anti-podal fixed points which are separated by a distance proportional to the momentum quadrature of the nanomechanical resonator. The difference between the second and third classes of fixed points is that one exists as a pair of fixed points on the unit circle for all $\ph>0$ (this is the second listed pair for positive detuning $\om>0$, and the third listed pair for negative detuning $\om<0$). This class of fixed points is typically stable when it exists (below we will be more specific), and we will call these fixed points the ``unit circle pair''. We can gain some physical insight as to what is going on here by recalling the definition of $\ph$ in \eqref{e:phdef}; we see that $\ph>0$ corresponds to
\beq
 \<{1+\xi\ka\pr}^2 - \<{\xi\ga\pr}^2 > 0 \,.
\eeq
This is simple in the case $\xi\ka\pr\gg1$ corresponding to a strong parametric drive (for $\xi$ fixed). Thus, the change from $\ph>0$ to $\ph<0$ corresponds to the mechanical damping dominating the parametric drive. The remaining class of the second and third classes (the third listed pair for positive detuning, and the second listed pair for negative detuning) only exists for $0<\ph<1$. At $\ph=0$ these fixed points appear (or annihilate for decreasing rather than increasing $\ph$) with the unit circle pair; and then head either towards the origin (for positive detuning) or out to infinity (for negative detuning) where they then disappear (or appear for decreasing rather than increasing $\ph$). This class is always unstable (again more on stability below), and we will call these fixed points the ``off-circle pair''. Physically, $0<\ph<1$ corresponds to either weak damping and weak parametric pumping, or, if they are held fixed, to $\xi$ being small and hence strong coupling.

\begin{figure}[!htbp]\begin{center}
\includegraphics[scale=0.75]{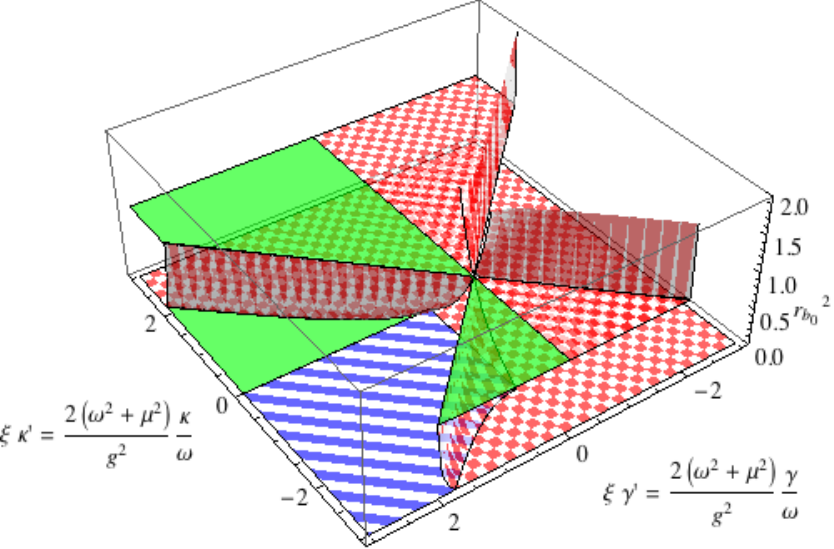}
\caption{(Color online) Radial components of the nanomechanical oscillator amplitude of the semi-classical fixed points. The existence and radial components of the semi-classical fixed points are functions of only one parameter $\ph=\<{1+\xi\ka\pr}^2 - \<{\xi\ga\pr}^2$, a particular combination of the scaled normalised parametric pumping $\xi\ka\pr$ and the scaled normalised nanomechanical dissipation rate $\xi\ga\pr$ (compare equation \eqref{e:scfp}). Here, the fixed points are plotted against the two component parameters for greater clarity, particularly in regards to stability. There are three different classes of fixed points: the origin ($r_{b_0}^2=0$) exists for all parameter values; the unit circle pair ($r_{b_0}^2=1$) exists for $\ph>0$; and the off-circle pair exists for $0<\ph<1$. The colours of the plot indicate the stability: green indicates stable fixed points; checkered red indicates unstable fixed points; and striped blue indicates fixed points that may or may not be stable. The stability is dependent on two additional parameters, a normalised microwave cavity dissipation $\mu\pr$, and normalised microwave cavity-nanomechanical coupling $\xi$, as the striped blue region requires. Note that since the sign of the microwave cavity detuning $\sgn\<{\om}=\sgn\<{\xi\ga\pr}$, the positive $\xi\ga\pr$-axis of the diagram represents positive detuning and the negative $\xi\ga\pr$-axis represents negative detuning. It is clear that negative detuning leads to semi-classical instability.}
\label{f:fp}
\end{center}\end{figure}

\begin{figure}[!htbp]\begin{center}
\includegraphics[scale=0.75]{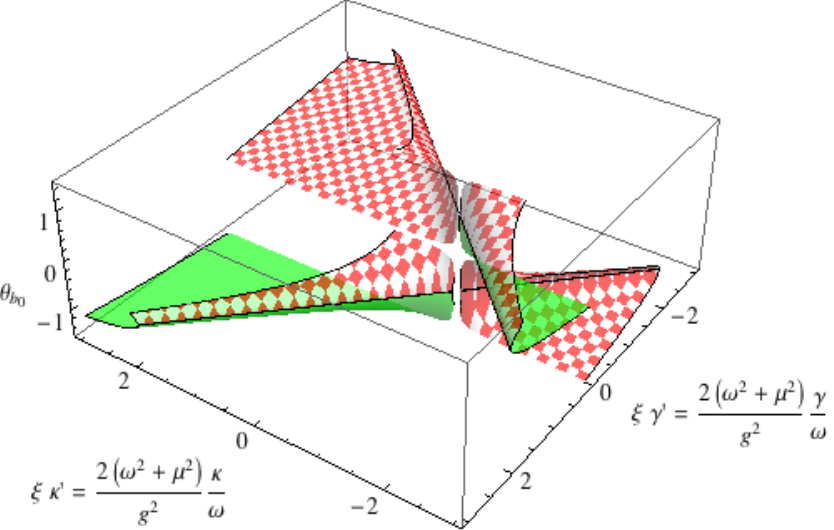}
\caption{(Color online) Angular components of the nanomechanical oscillator amplitude of the semi-classical fixed points (compare equation \eqref{e:scfp}). The fixed point at origin is not plotted for the obvious reason that its angular component is undefined. The angular components of the unit circle and off-circle pairs of semi-classical fixed points (the two classes of fixed points other than the origin) are functions of two parameters, the scaled normalised parametric pumping $\xi\ka\pr$ and the scaled normalised nanomechanical dissipation rate $\xi\ga\pr$. The angular components of the fixed points are plotted against these. The colours of the plot indicate the stability: green indicates stable fixed points; and checkered red indicates unstable fixed points. The unit circle pair are the stable fixed points in the positive $\xi\ga\pr$ (positive detuning) half-plane. It is seen that for values of the scaled normalised parametric pumping $\xi\ka\pr>-1$, the unit circle pair tends to be increasingly aligned with the y-axis. For values of the scaled normalised parametric pumping $\xi\ka\pr<-1$, the unit circle pair tends to be increasingly aligned with the x-axis. Note that the apparent discontinuity in crossing the $\xi\ka\pr$-axis is not real, since to cross means a flipping of the sign $\xi\ga\pr$ and the detuning $\om$ (which causes a pole in the parameters as the detuning crosses $0$), and which also flips the sign of $\xi\ka\pr$. To make a clean crossing and keep the sign of $\xi\ka\pr$, the sign of the parametric pumping must be flipped together with the detuning. It is best to consider positive and negative detuning (positive and negative values of the normalised scaled nanomechanical dissipation rate $\xi\ga\pr$) separately.}
\label{f:fpa}
\end{center}\end{figure}

The important observation to make at this point is that the existence and the positions of the fixed points depend upon only two parameters, $\xi\ka\pr$ and $\xi\ga\pr$ (since $\ph$ is a function of only these two parameters, and $\sgn\<{\om}=\sgn\<{\xi\ga\pr}$). In fact, the existence of the fixed points depends only upon the single parameter $\ph$. Additionally, one reason why this particular parametrisation was chosen is now quite clear --- it places the possibly stable pair of fixed points on the unit circle. We plot the radial and angular components of the nanomechanical components of the fixed points in figures \ref{f:fp} and \ref{f:fpa} respectively. The colours in these figures show the stability which is now discussed.

We also note that $\aop\to-\aop$, $\bop\to-\bop$ is a symmetry of the system, and thus pairs of anti-podal fixed points are the expected semi-classical result in both factor spaces.


\begin{figure}[!htbp]\begin{center}
\includegraphics[scale=1]{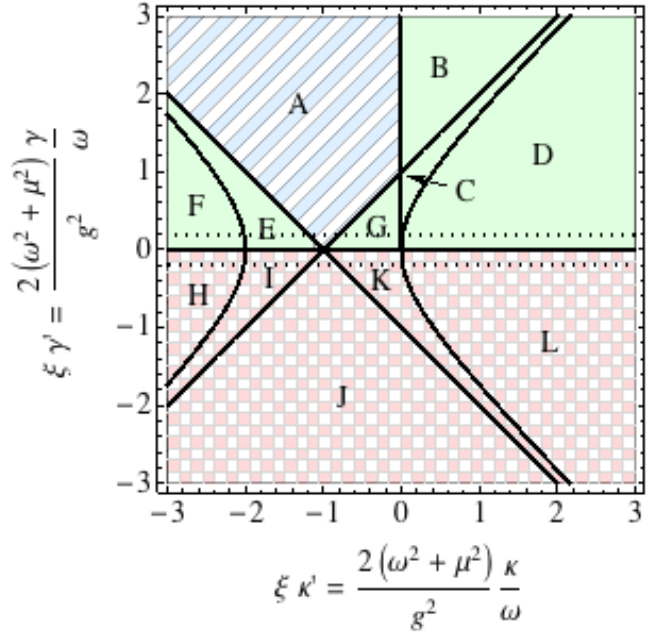}
\caption{(Color online) A two-dimensional projection of the ``phase diagram'' of the semi-classical microwave cavity-nanomechanical system. The qualitatively different regions of semi-classical fixed point structures are plotted against the scaled normalised parametric pumping $\xi\ka\pr$ and the scaled normalised nanomechanical dissipation rate $\xi\ga\pr$. Stability depends upon these two parameters and also an additional two: the normalised microwave cavity dissipation rate $\mu\pr$, and the normalised microwave cavity-nanomechanical coupling $\xi$. The existence and positions of the fixed points depend upon only the first two parameters, making this two-dimensional projection quite useful. The colours of the regions indicate the stability: green indicates the region contains some stable fixed points; checkered red indicates the region contains only unstable fixed points; and striped blue indicates the region may or may not contain a stable fixed point (depending on the additional two parameters). Note that since the sign of microwave cavity detuning $\sgn\<{\om}=\sgn\<{\xi\ga\pr}$, the top half of the diagram represents positive detuning and the bottom half of the diagram represents negative detuning. It is clear that negative detuning leads to semi-classical instability. The origin is always a semi-classical fixed point. It is stable in regions B and G; and may be stable or unstable, including undergoing Hopf bifurcations, in regions A, E, and G. The fixed point pair on the nanomechanical unit circle exists in regions C, D, E, F, G, H, I, K, and L; and is stable in regions C, D, E, F, and G. The off-circle fixed points exist in regions C, E, G, I, K; but are always unstable. A fuller explanation of the semi-classical dynamics is given in the text. The upper dotted line shows the location in parameter space of the systems shown in Figures \ref{f:hscp}, \ref{f:hqp}, and \ref{f:hqpc}; the lower dotted line shows the location in parameter space of the systems shown in Figures \ref{f:hscn}, \ref{f:hqn}, and \ref{f:hqnc}. The lines are close to the x-axis for ease of quantum numerical simulation as explained in Section \ref{s:q}.}
\label{f:bif}
\end{center}\end{figure}

\begin{figure}[!htbp]\begin{center}
\includegraphics[scale=1]{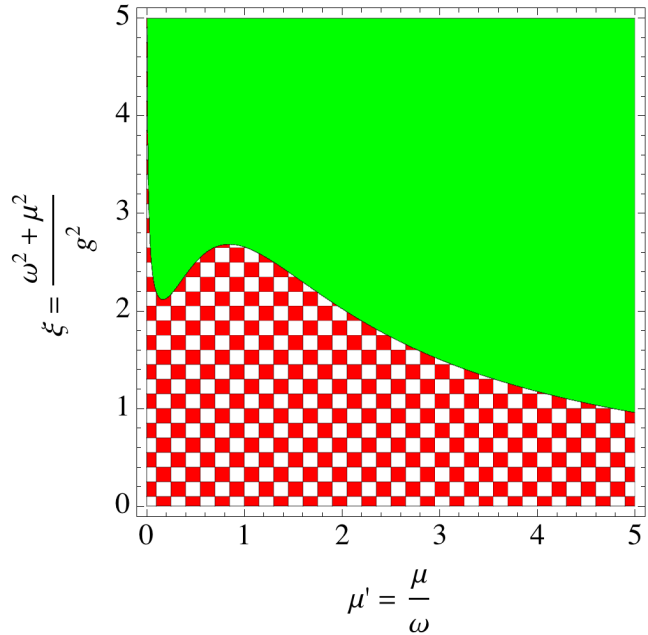}
\caption{(Color online) A two-dimensional ``phase diagram'' of the semi-classical microwave cavity-nanomechanical system at a particular value of the scaled normalised parametric pumping $\xi\ka\pr=-0.8$ and the scaled normalised mechanical dissipation rate $\xi\ga\pr=0.3$. This phase diagram thus shows the extra two dimensions present at a particular point in region A in the two-dimensional projection of Figure \ref{f:bif} of the full four-dimensional phase space. These extra two dimensions are the normalised microwave cavity dissipation rate $\mu\pr$, and the normalised microwave cavity-nanomechanics coupling $\xi$. The colours of the regions indicate the stability of the fixed point at the origin: green indicates the origin is stable; checkered red indicates the origin is unstable. For large values of either the mechanical dissipation rate or the normalised coupling, the origin is generally stable.}
\label{f:hb}
\end{center}\end{figure}

The stability of the fixed points is not quite as simple. We see that although the existence and the positions of fixed points depend upon only two parameters ($\xi\ka\pr$ and $\xi\ga\pr$), the stability is dependent on four: $\xi\ka\pr$ and $\xi\ga\pr$; but also $\mu\pr$ and $\xi$. Thus, a bifurcation or stability diagram must be four-dimensional. Nonetheless, we can project upon to the two-dimensional space spanned by $\xi\ka\pr$ and $\xi\ga\pr$ to discuss the bifurcation structure somewhat effectively. Importantly, to investigate all of the potential switching of stabilities, a large number of trials of random sampling in the four-dimensional parameter space was used in addition to analytic methods. This numeric method used a logarithmic distribution spanning $7$ orders of magnitude in each parameter. Where either the stability was analytically shown to be stable or unstable, or the numeric method yielded $>99\%$ of trials as either stable or unstable, we have indicated the stability of the parameter space as stable or unstable respectively. The resulting two-dimensional diagram is plotted in Figure \ref{f:bif}. Later, we show an example of a mixed region with stable and unstable regions depending on all four parameters in Figure \ref{f:hb}.

Referring to the ``phase diagram'' of Figure \ref{f:bif}, we discuss the various regions (or ``phases'') of the semi-classical dynamical system. The plot of the fixed points themselves on the same axes is also helpful, and the reader is encouraged to also refer to Figure \ref{f:fp} during the discussion. The first point to note is that the upper half-plane indicates positive detuning ($\om>0$), and that the lower half-plane indicates negative detuning ($\om<0$). From the diagram it is clear that negative detuning causes widespread instability of the semi-classical system. Positive detuning is thus a better candidate for comparison to the full quantum steady state of the next section, though both will be compared.

We first introduce some dynamical systems terminology for the bifurcations occurring along region boundaries. Briefly, a pitchfork bifurcation consists of one fixed point changing stability and giving birth to two new fixed points. The pitchfork bifurcation is supercritical if two stable fixed points are born when one stable fixed point loses stability, and is subcritical if two unstable fixed points are born when one unstable fixed point gains stability. Saddle node bifurcations consist of the annihilation of two fixed points, leaving none, or alternatively, the creation of two fixed points where there were previously none. Also present are Hopf bifurcations of the fixed point at the origin. A Hopf bifurcation consists of a fixed point changing stability and giving birth to a limit cycle. A Hopf bifurcation is supercritical if a stable fixed point loses stability and gives birth to a stable limit cycle, and is subcritical if an unstable fixed point gains stability giving birth to an unstable limit cycle. For more information about bifurcations in dynamical systems, see for example \cite{glendinning:1994}.

For positive detuning, starting at a large negative parametric pumping (and hence a negative value of $\xi\ka\pr$), as we move from region F through regions E, A, G (or B), and C, to D, there is a marked qualitative change in the semi-classical steady state. Ignoring for the moment the changes in stability of the origin, there are two fixed points on the unit circle which are initially aligned with the x-axis, these gradually rotate before being annihilated at the E-A boundary (in a saddle-node bifurcation with an unstable pair) before reappearing at the A-G (or B-C) boundary (again in a saddle-node bifurcation with an unstable pair) and then gradually rotating towards being aligned with the y-axis. The unstable pair with which the annihilation of the unit circle pair occurs, is born and dies at the origin in a subcritical pitchfork bifurcation. For negative detuning, starting at a large negative parametric pumping (and hence a positive value of $\xi\ka\pr$, since $\sgn\<{\xi\ka\pr}=\sgn{\om}\sgn{\ka}$), as we move from region L through regions K, J, and I, to H, there is a similar change in the semi-classical steady state. However, all of the negative detuned fixed points are unstable, so the similarity is only in terms of the fixed point positions. Again ignoring the origin, the two fixed points on the unit circle are initially aligned with the x-axis, these gradually rotate before being annihilated at the K-J boundary before reappearing at the J-I boundary and then gradually rotating towards being aligned with the y-axis. We will seek to compare this movement of the fixed points with the changes in the quantum steady state phase space in the next section.

\begin{figure}[!htbp]\begin{center}
\includegraphics[scale=0.8]{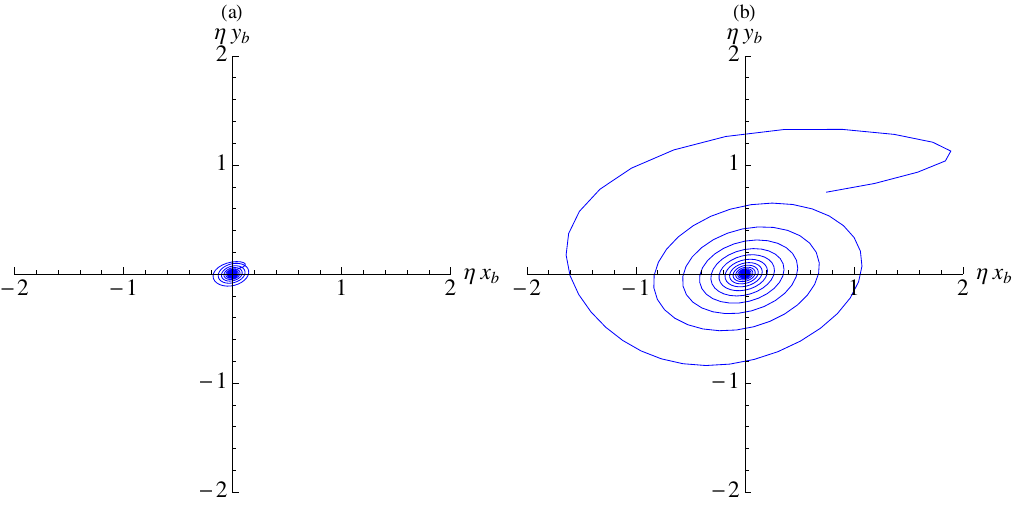}
\caption{(Color online) The nanomechanical components of a trajectory through the semi-classical phase space on the stable side of a Hopf bifurcation of the origin, showing a stable attractive fixed point at the origin. The parameters for this trajectory are: a scaled normalised parametric pumping of $\xi\ka\pr=-0.8$; a scaled normalised mechanical dissipation rate of $\xi\ga\pr=0.3$; a normalised microwave cavity dissipation rate of $\mu\pr=0.3$; and a normalised microwave cavity-nanomechanics coupling of $\xi=2.4$. This places the system in region region A of Figure \ref{f:bif}, and the green region of Figure \ref{f:hb}. Qualitatively similar behaviour is found for other system parameter values lying in these same two regions; contrast with Figure \ref{f:hut}. The trajectory of plot (a) starts close the origin at $x_{a_0}=y_{a_0}=x_{b_0}=y_{b_0}=0.05$. The trajectory of plot (b) starts away from the origin at $x_{a_0}=y_{a_0}=x_{b_0}=y_{b_0}=0.75$. Both trajectories are seen to be attracted to the stable fixed point at the origin. Backing out the original couplings necessary for these example parameter values, this trajectory is for $\om=3,\ch=0.1,\ka=-1,g=2.86,\mu=0.9,\ga=0.38$.}
\label{f:hst}
\end{center}\end{figure}

\begin{figure}[!htbp]\begin{center}
\includegraphics[scale=0.8]{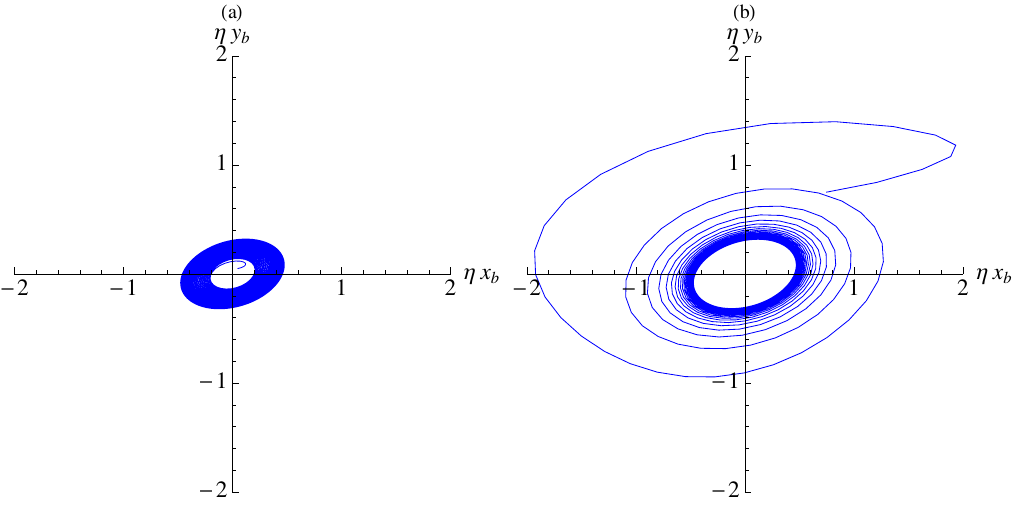}
\caption{(Color online) The nanomechanical components of a trajectory through the semi-classical phase space on the unstable side of a Hopf bifurcation of the origin, showing a stable attractive limit cycle. The parameters for this trajectory are: a scaled normalised parametric pumping of $\xi\ka\pr=-0.8$; a scaled normalised mechanical dissipation rate of $\xi\ga\pr=0.3$; a normalised microwave cavity dissipation rate of $\mu\pr=0.3$; and a normalised microwave cavity-nanomechanics coupling of $\xi=2.2$. This places the system in region region A of Figure \ref{f:bif}, and the checkered red region of Figure \ref{f:hb}. Qualitatively similar behaviour is found for other system parameter values lying in these same two regions; contrast with Figure \ref{f:hst}. The trajectory of plot (a) starts close the origin at $x_{a_0}=y_{a_0}=x_{b_0}=y_{b_0}=0.05$. The trajectory of plot (b) starts away from the origin at $x_{a_0}=y_{a_0}=x_{b_0}=y_{b_0}=0.75$. Trajectory (a) shows that the fixed point at the origin has become unstable, and trajectories (a) and (b) together that a stable attracting limit cycle has formed. Backing out the original couplings necessary for these example parameter values, this trajectory is for $\om=3,\ch=0.1,\ka=-1.09,g=2.99,\mu=0.9,\ga=0.4$.}
\label{f:hut}
\end{center}\end{figure}

Returning to the changes in stability of the origin, there are potential Hopf bifurcations \cite{glendinning:1994} in regions A, E, and G, dependent on the additional two stability parameters, the normalised microwave cavity dissipation parameter $\mu\pr$ and the normalised coupling parameter $\xi$. The Hopf bifurcation causes the stable origin to become unstable, and creates a stable limit cycle surrounding it. In order to show this behaviour, we show trajectories of the semi-classical system for two different values of $\mu\pr$ and $\xi$ for a particular point in region A. Figure \ref{f:hb} shows the stability at a particular point in region A as a function of the two additional stability parameters. Figure \ref{f:hst} shows two trajectories on the stable side of the Hopf bifurcation, while Figure \ref{f:hut} shows two trajectories on the unstable side of the Hopf bifurcation. We see from figures \ref{f:hst} and \ref{f:hut} that the origin switches from attracting nearby trajectories to repelling them, and that an attracting limit cycle is born as the origin loses its stability. In the next section we will see what happens to the quantum steady state phase space as we move through this semi-classical bifurcation.


\begin{figure}[!htbp]\begin{center}
\includegraphics[scale=0.6]{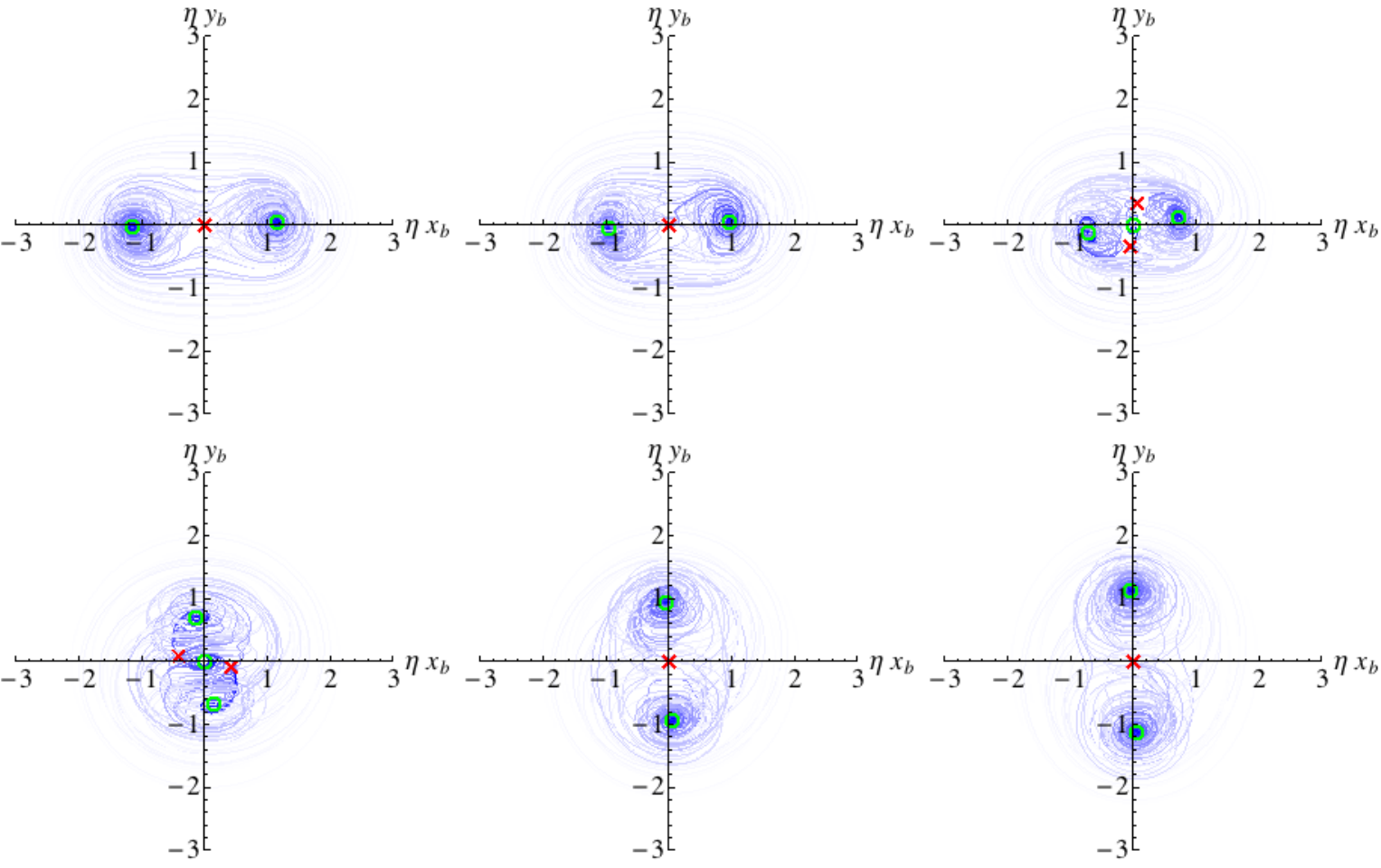}
\caption{(Color online) Some semi-classical trajectories for positive detuning ($\om>0$) through a linear sweep of increasing parametric pumping. The stable semi-classical fixed points are marked with a green circle, and the unstable semi-classical fixed points are marked with a red circle. The trajectories start as white/light blue and move towards blue, such that the darker blue is the better approximation of the steady state. The axes here are scaled as the original parameters, so that they can be compared to the quantum Wigner functions in Figure \ref{f:hqp}. The trajectories are clearly attracted to the stable semi-classical fixed points. Backing out the original couplings necessary for the example parameter values plotted for, $\om=2,\ch=1,g=1.16,\mu=0.2,\ga=0.07$. The parametric pumping $\ka$ is linearly swept from $\ka=-8$ to $\ka=4$ in the six plots (along the upper dotted line in Figure \ref{f:bif}). There is qualitatively similar behaviour for all systems in the same region of the phase space diagram Figure \ref{f:bif}. In terms of the parametrisation used in the text, and in the phase diagram of Figure \ref{f:bif}, we have $\xi\ga\pr=0.2$ while $\xi\ka\pr$ is swept from $\xi\ka\pr=-4$ to $\xi\ka\pr=2$.}
\label{f:hscp}
\end{center}\end{figure}

\begin{figure}[!htbp]\begin{center}
\includegraphics[scale=0.6]{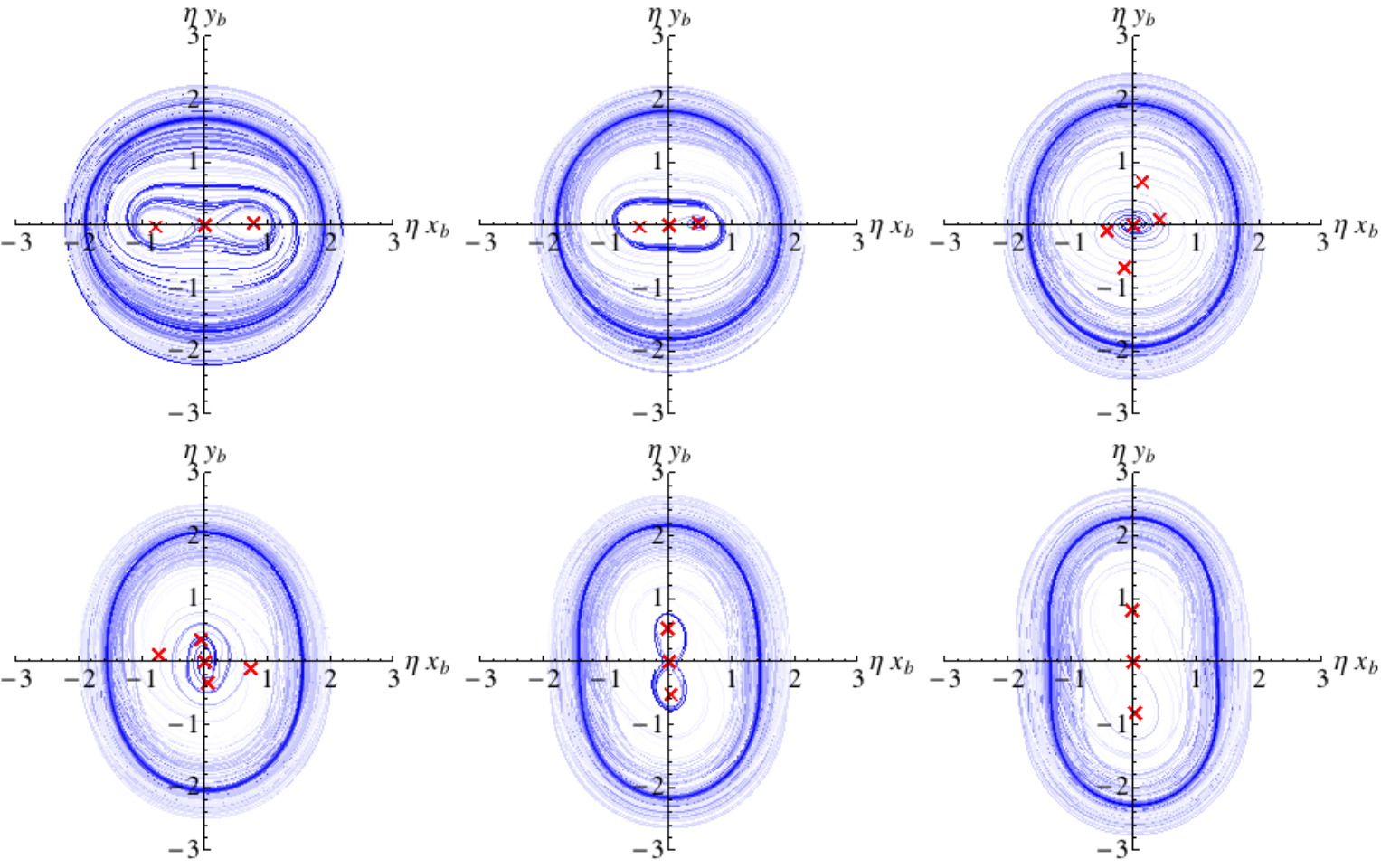}
\caption{(Color online) Some semi-classical trajectories for negative detuning ($\om<0$) through a linear sweep of increasing parametric pumping. The stable semi-classical fixed points are marked with a green circle, and the unstable semi-classical fixed points are marked with a red circle. The trajectories start as white/light blue and move towards blue, such that the darker blue is the better approximation of the steady state. The axes here are scaled as the original parameters, so that they can be compared to the quantum Wigner functions in Figure \ref{f:hqn}. The trajectories are clearly repelled by the unstable semi-classical fixed points, out to an outer limit cycle. Backing out the original couplings necessary for the example parameter values plotted for, $\om=-2,\ch=1,g=1.16,\mu=0.2,\ga=0.07$. The parametric pumping $\ka$ is linearly swept from $\ka=-8$ to $\ka=4$ in the six plots (along the lower dotted line in Figure \ref{f:bif}). There is qualitatively similar behaviour for all systems in the same region of the phase space diagram Figure \ref{f:bif}. In terms of the parametrisation used in the text, and in the phase diagram of Figure \ref{f:bif}, we have $\xi\ga\pr=-0.2$ while $\xi\ka\pr$ is swept from $\xi\ka\pr=2$ to $\xi\ka\pr=-4$.}
\label{f:hscn}
\end{center}\end{figure}

Lastly, to demonstrate the effect of the unstable fixed points, and in particular to attempt to get a semi-classical picture for the negative detuning steady state, we investigate some trajectories through phase space. Figure \ref{f:hscp} shows some trajectories for a linear sweep of increasing parametric pumping $\ka$ (and thus increasing $\xi\ka\pr$) with all other couplings held constant. Figure \ref{f:hscn} shows some trajectories for the same linear sweep, with all the same couplings, with the exception of a switched sign for the detuning. We note the positive detuning trajectories being attracted to the stable semi-classical fixed points, and the negative detuning trajectories heading away from the unstable semi-classical fixed points and out into surrounding limit cycles.


\section{Quantum steady states.}

\label{s:q}

\subsection{Quantum steady state phase space.}

In the previous section we showed the fixed point bifurcations of the semi-classical system. Here, we investigate whether there is a signature of the semi-classical bifurcations present in the full quantum description. We do this by numerically computing the quantum steady state density operator, and a corresponding Wigner function, in different regions of the semi-classical ``phase diagram''. It is hoped that by sweeping the coupling parameters through a semi-classical bifurcation, there will be a corresponding qualitative change in the quantum steady state manifest in the corresponding Wigner function in phase space. This kind of correspondence principle has proven to be the case for other dissipative nonlinear quantum systems \cite{hines:2004,meaney:2010,Furuya,Santhanam,Nemes,Brandes}.

To perform the numerical computation of the quantum steady state we use the Quantum Optics MATLAB toolbox \cite{tan:2002}. To do this we approximate the infinite basis of each oscillator by truncating in the Fock (number) basis at $14\approx3.74^2$. This means that we must choose couplings such that the bifurcation takes place sufficiently close to the origin to be accurately approximated by the truncation. This is roughly because a coherent state of amplitude $\al$ has a mean occupation number of $\abs{\al}^2$. Given the quantum steady state typically (as we shall see direct evidence of in this section) has support centred on the semi-classical steady state, fixed points far from the origin (high $\abs{\al}$) will produce high occupations and thus inaccurate results if we truncate in the Fock (number) basis.

\begin{figure}[!htbp]\begin{center}
\includegraphics[scale=0.6]{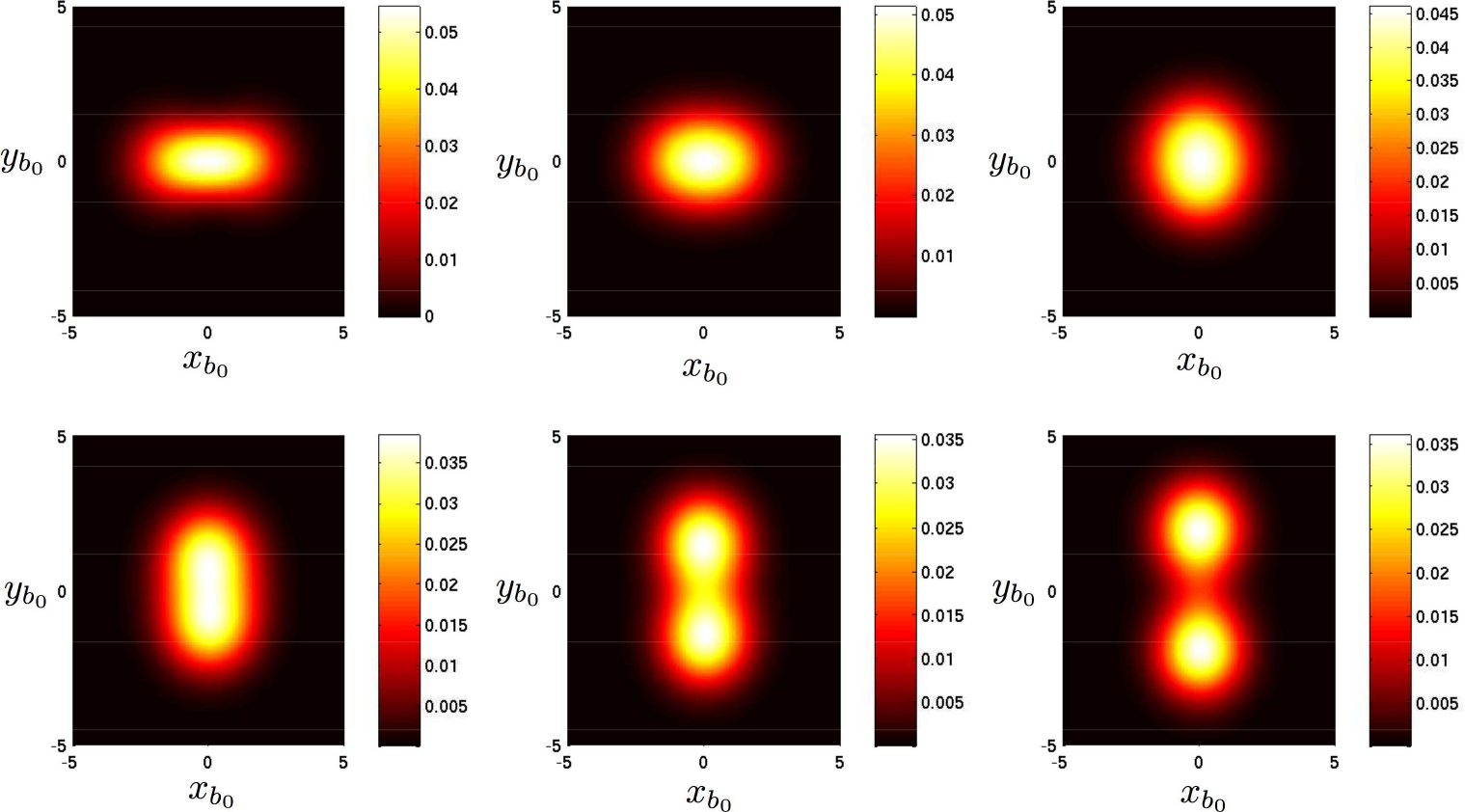}
\caption{(Color online) Density plots of Wigner functions of the nano-mechanical factor space for positive detuning ($\om>0$) through a linear sweep of increasing parametric pumping. The Wigner function $W(x,y)$ is plotted where $x$ and $y$ are the position and momentum quadratures (though not scaled as such) of the nanomechanical oscillator. The quantum steady state shows clear signs of the semi-classical bifurcations it undergoes. Particular comparison can be made to the semi-classical trajectories of Figure \ref{f:hscp}. Whilst there is evidence of the semi-classical bifurcations, the quantum fluctuations wash out the sharp transitions. In particular, the semi-classical bifurcations between regions E, A, G, and C of the semi-classical phase diagram \ref{f:bif} are not discenible. However, the signature of the overall transition from x-axis aligned density in region F to y-axis aligned density in region D is clearly visible. Backing out the original couplings necessary for the example parameter values plotted for, $\om=2,\ch=1,g=1.16,\mu=0.2,\ga=0.07$. The parametric pumping $\ka$ is linearly swept from $\ka=-8$ to $\ka=4$ in the six plots (along the upper dotted line in Figure \ref{f:bif}). In terms of the parametrisation used in the text, and in the phase diagram of Figure \ref{f:bif}, we have $\xi\ga\pr=0.2$ while $\xi\ka\pr$ is swept from $\xi\ka\pr=-4$ to $\xi\ka\pr=2$.}
\label{f:hqp}
\end{center}\end{figure}

\begin{figure}[!htbp]\begin{center}
\includegraphics[scale=0.6]{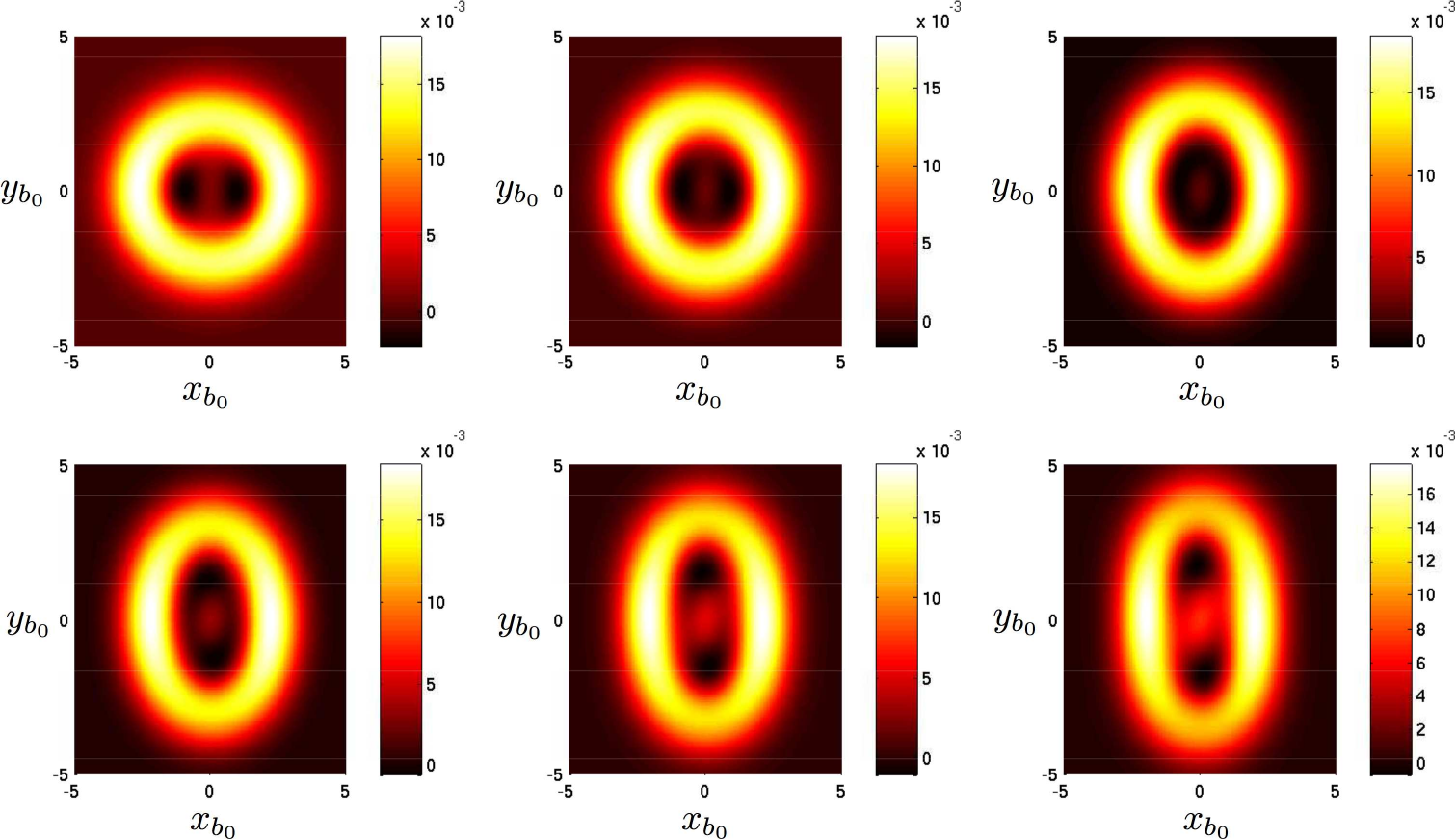}
\caption{(Color online) Density plots of Wigner functions of the nano-mechanical factor space for negative detuning ($\om<0$) through a linear sweep of increasing parametric pumping. The Wigner function $W(x,y)$ is plotted where $x$ and $y$ are the position and momentum quadratures (though not scaled as such) of the nanomechanical oscillator. Particular comparison can be made to the semi-classical trajectories of Figure \ref{f:hscn}. The major and minor axes of symmetry of the phase-diffused rings are seen to be the same as the semi-classical limit cycles. Backing out the original couplings necessary for the example parameter values plotted for, $\om=-2,\ch=1,g=1.16,\mu=0.2,\ga=0.07$. The parametric pumping $\ka$ is linearly swept from $\ka=-8$ to $\ka=4$ in the six plots (along the lower dotted line in Figure \ref{f:bif}). In terms of the parametrisation used in the text, and in the phase diagram of Figure \ref{f:bif}, we have $\xi\ga\pr=-0.2$ while $\xi\ka\pr$ is swept from $\xi\ka\pr=2$ to $\xi\ka\pr=-4$.}
\label{f:hqn}
\end{center}\end{figure}

\begin{figure}[!htbp]\begin{center}
\includegraphics[scale=0.6]{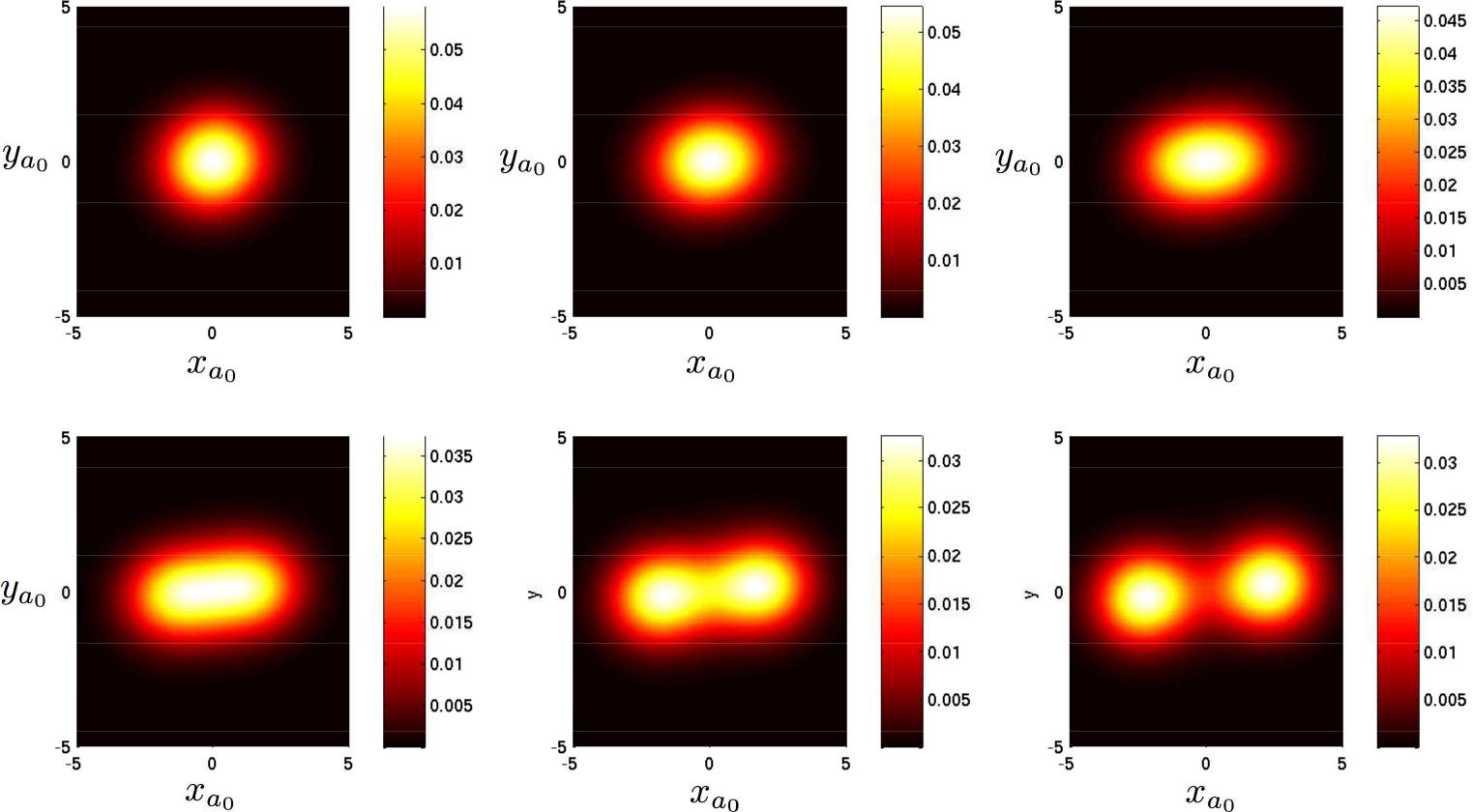}
\caption{(Color online) Density plots of Wigner functions of the superconducting microwave cavity factor space for positive detuning ($\om>0$) through a linear sweep of increasing parametric pumping. The Wigner function $W(x,y)$ is plotted where $x$ and $y$ are two quadratures operators of the superconducting microwave cavity field. The quantum steady state shows clear signs of the semi-classical bifurcations it undergoes. Backing out the original couplings necessary for the example parameter values plotted for, $\om=2,\ch=1,g=1.16,\mu=0.2,\ga=0.07$. The parametric pumping $\ka$ is linearly swept from $\ka=-8$ to $\ka=4$ in the six plots (along the upper dotted line in Figure \ref{f:bif}). In terms of the parametrisation used in the text, and in the phase diagram of Figure \ref{f:bif}, we have $\xi\ga\pr=0.2$ while $\xi\ka\pr$ is swept from $\xi\ka\pr=-4$ to $\xi\ka\pr=2$.}
\label{f:hqpc}
\end{center}\end{figure}

\begin{figure}[!htbp]\begin{center}
\includegraphics[scale=0.6]{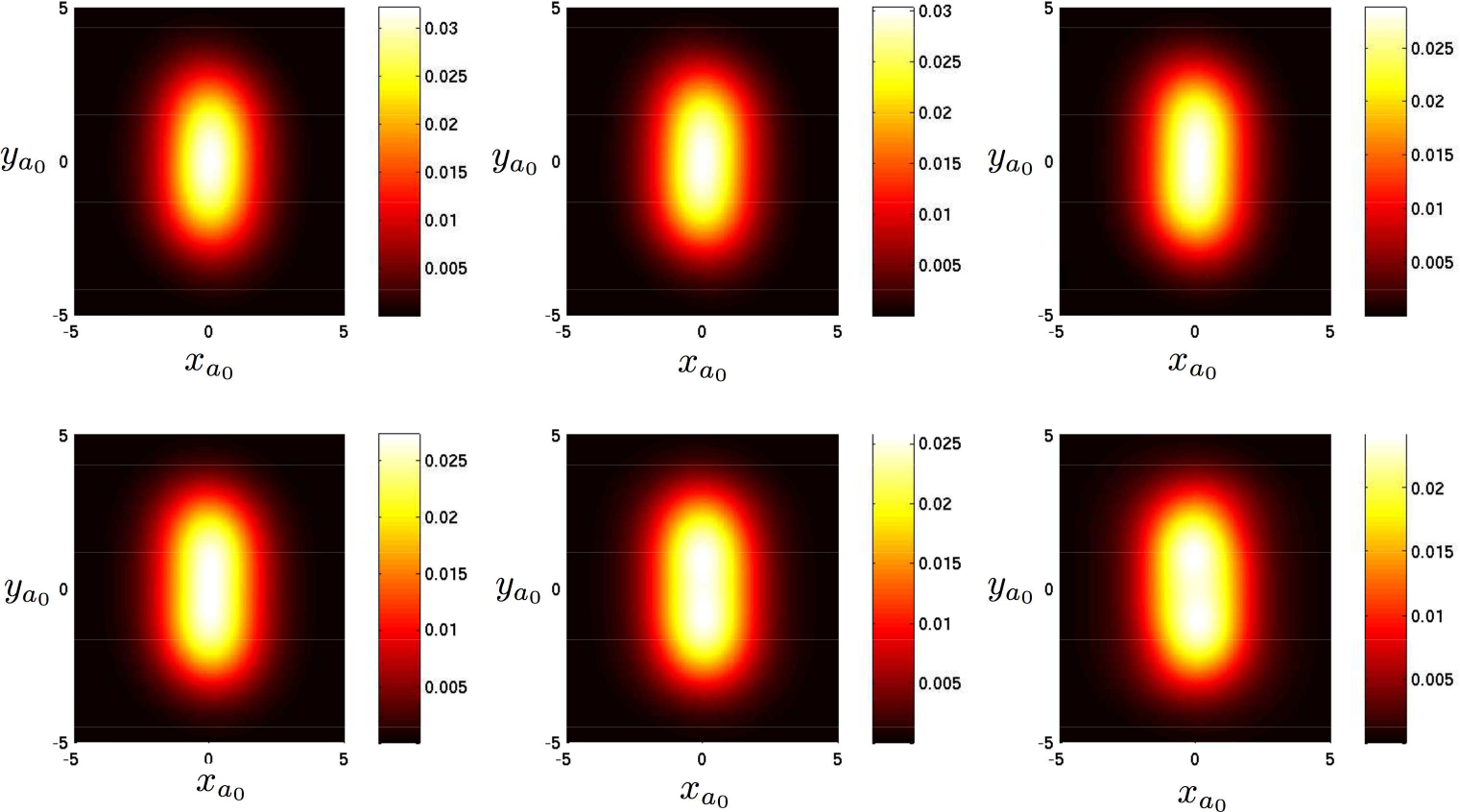}
\caption{(Color online) Density plots of Wigner functions of the superconducting microwave cavity factor space for negative detuning ($\om<0$) through a linear sweep of increasing parametric pumping. The Wigner function $W(x,y)$ is plotted where $x$ and $y$ are two quadratures operators of the superconducting microwave cavity field. Backing out the original couplings necessary for the example parameter values plotted for, $\om=-2,\ch=1,g=1.16,\mu=0.2,\ga=0.07$. The parametric pumping $\ka$ is linearly swept from $\ka=-8$ to $\ka=4$ in the six plots (along the lower dotted line in Figure \ref{f:bif}). In terms of the parametrisation used in the text, and in the phase diagram of Figure \ref{f:bif}, we have $\xi\ga\pr=-0.2$ while $\xi\ka\pr$ is swept from $\xi\ka\pr=2$ to $\xi\ka\pr=-4$.}
\label{f:hqnc}
\end{center}\end{figure}

We hold all couplings equal save the parametric pumping $\ka$ (and thus $\xi\ka\pr$), and move along from left to right in the upper half plane of the phase diagram of Figure \ref{f:bif}. The couplings we choose are the same as those we chose for the semi-classical trajectories in the previous section to allow for direct comparison. These were $\om=\pm2,\ch=1,g=1.16,\mu=0.2,\ga=0.07$. The parametric pumping $\ka$ is linearly swept from $\ka=-8$ to $\ka=4$ in the six plots that follow for both positive ($\om>0$) and negative ($\om<0$) detuning. In terms of the parametrisation used in the text, and in the phase diagram of Figure \ref{f:bif}, we have $\xi\ga\pr=-0.2$ while $\xi\ka\pr$ is swept from $\xi\ka\pr=-4$ to $\xi\ka\pr=2$. To visualise the microwave cavity and nanomechanical factor spaces we choose to look at their phase spaces by using the Wigner function of the reduced density matrix for each system. The Wigner function is defined as $W\<{x,y}=\f{\pi\hb}\int_{-\infty}^\infty\dz\BOK{x-z}{\dm}{x+z}\ee^{\ii\f[2yz]{\hb}}$; for more on the Wigner function see \cite{zachos:2005,walls:2007}. We plot the series of Wigner functions of the nano-mechanical factor space in Figure \ref{f:hqp} for positive detuning and in Figure \ref{f:hqn} for negative detuning. For completeness, we also show the corresponding Wigner functions for the microwave cavity factor space in figures \ref{f:hqpc} and \ref{f:hqnc} for positive and negative detuning respectively.


For positive detuning, there is a clear signature of the semi-classical bifurcation. For negative detuning, the semi-classical fixed points are unstable, though the limit cycle apparent in the trajectories of Figure \ref{f:hscn} are a signature which appears to be present in the quantum steady state phase space as seen in Figure \ref{f:hqn}. Certainly, the semi-classical approximation used in the previous section provides a useful heuristic for understanding the correlations inherent in quantum steady states. 

\subsection{Quantum steady state entanglement.}

We wish to determine the steady state quantum entanglement between the superconducting microwave resonator and the nonlinear nanomechanical resonator. The complete system consists of three parts: the superconducting microwave resonator; the nanomechanical resonator; and the environment. A natural measure of such entanglement is the log negativity \cite{vidal:2002}. With access to the truncated steady state density matrix, we can assume the truncated matrix is a sufficiently good approximation and directly compute the log negativity $E_N$ as the base $2$ logarithm of the trace norm of the partial transpose of the bipartite density matrix,
\beq
 E_N\<{\dm} = \log_2\norm[1]{\te{\rh_0}^{T_A}} \,.
\eeq
We compute the log negativity for each quantum steady state in a linear sweep of the parametric pumping $\ka$ whilst holding other parameters constant at the same values used to investigate the quantum steady state phase space in the previous section. We note that the entanglement entropy of similar nonlinear systems has been examined, though with the quartic nonlinearity treated as a perturbation \cite{vidal:2007}.

\begin{figure}[!htbp]\begin{center}
\includegraphics[scale=0.7]{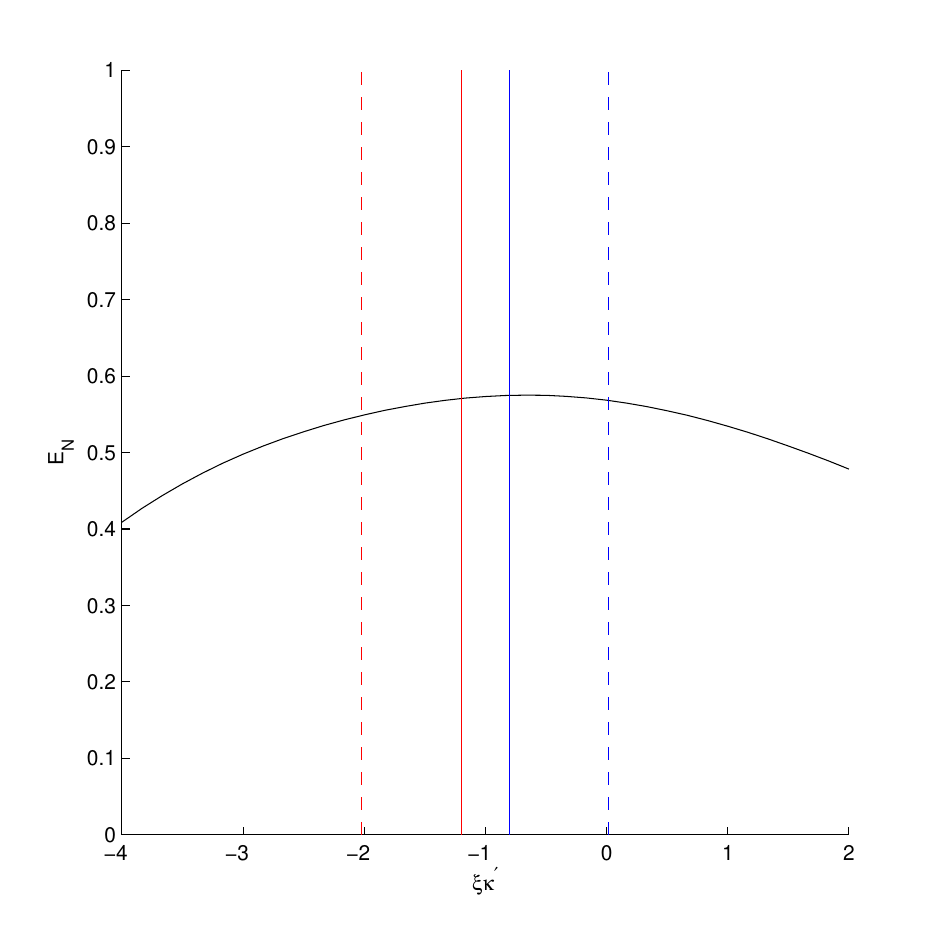}
\caption{(Color online) Quantum entanglement, as measured by the logarithmic negativity, for positive detuning ($\om>0$) through a linear sweep of increasing parametric pumping. The four vertical lines show the positions of the semi-classical bifurcations: the first vertical line (the dotted red line) is where the two unstable off-circle fixed points appear at the origin (the F-E boundary); the second vertical line (the solid red line) is where just created unstable fixed points annihilate the unit circle pair (the E-A boundary); the third vertical line (the solid blue line) is where the two pairs of fixed points re-appear (the A-G boundary); and the fourth vertical line (the dotted blue line) is where the off-circle pair vanishes at the origin (the C-D boundary). The quantum entanglement is peaked just to one side of the semi-classical bifurcation centre. Backing out the original couplings necessary for the example parameter values plotted for, $\om=2,\ch=1,g=1.16,\mu=0.2,\ga=0.07$. The parametric pumping $\ka$ is linearly swept from $\ka=-8$ to $\ka=4$. In terms of the parametrisation used in the text, and in the phase diagram of Figure \ref{f:bif}, we have $\xi\ga\pr=0.2$ while $\xi\ka\pr$ is swept from $\xi\ka\pr=-4$ to $\xi\ka\pr=2$.}
\label{f:ep}
\end{center}\end{figure}

\begin{figure}[!htbp]\begin{center}
\includegraphics[scale=0.7]{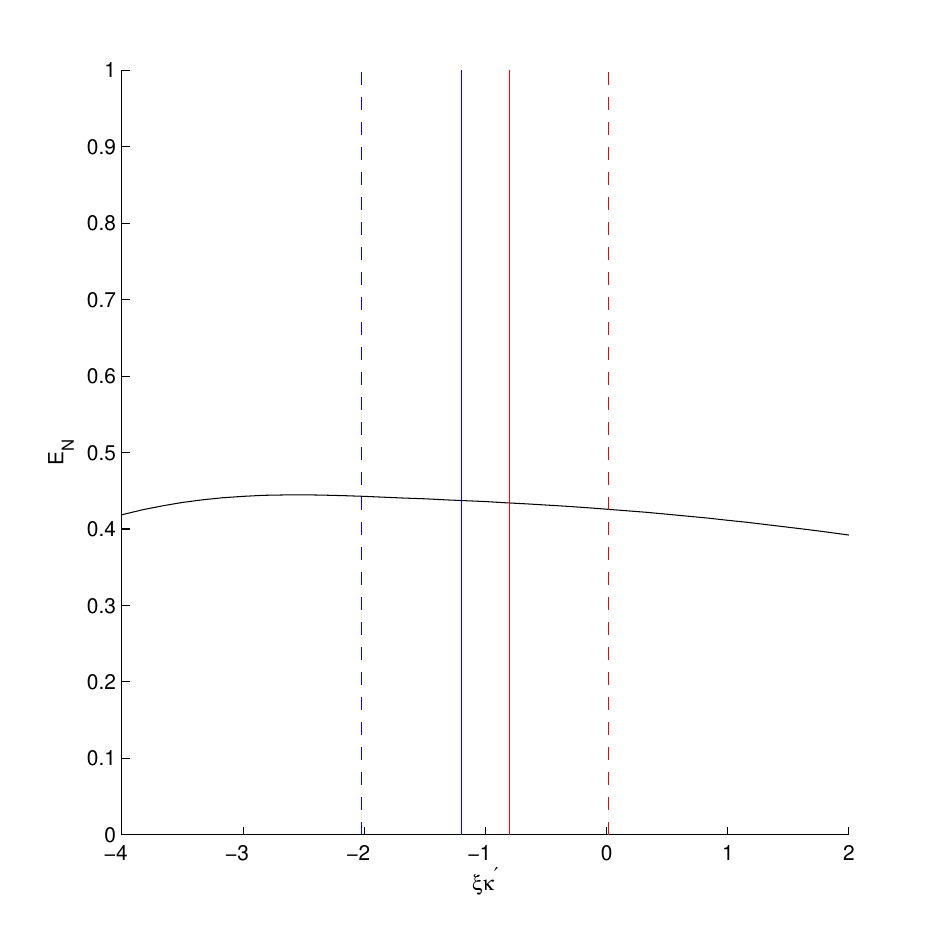}
\caption{(Color online) Quantum entanglement, as measured by the logarithmic negativity, for negative detuning ($\om<0$) through a linear sweep of increasing parametric pumping. The four vertical lines show the positions of the semi-classical bifurcations (from right to left): the first vertical line (the dotted red line) is where the two off-circle fixed points appear (the L-K boundary); the second vertical line (the solid red line) is where just created fixed points annihilate the unit circle pair (the K-J boundary); the third vertical line (the solid blue line) is where the two pairs of fixed points re-appear (the J-I boundary); and the fourth vertical line (the dotted blue line) is where the off-circle pair vanishes (the I-H boundary). The quantum entanglement for negative detuning is remarkably flat. Backing out the original couplings necessary for the example parameter values plotted for, $\om=-2,\ch=1,g=1.16,\mu=0.2,\ga=0.07$. The parametric pumping $\ka$ is linearly swept from $\ka=-8$ to $\ka=4$. In terms of the parametrisation used in the text, and in the phase diagram of Figure \ref{f:bif}, we have $\xi\ga\pr=-0.2$ while $\xi\ka\pr$ is swept from $\xi\ka\pr=2$ to $\xi\ka\pr=-4$.}
\label{f:en}
\end{center}\end{figure}

The quantum entanglement for the positive and negative detuning cases are plotted in figures \ref{f:ep} and \ref{f:en} respectively. In what we consider the more meaningful ``phase transition'', the bifurcation traversals in the positive detuning case of Figure \ref{f:ep}, we see that the entanglement is peaked slightly to one side of the bifurcation centre. Generally speaking, a system has more entanglement at criticality; however the exact peak can depend upon the entanglement measure. For example, using concurrence to measure entanglement for the transverse Ising model shows a peak just off criticality \cite{osborne:2002}; while Vidal et al have shown that entanglement entropy scaling with block size shows that for a sufficiently large, fixed block size, the entanglement entropy peak is always at criticality \cite{vidal:2003}. The fact that the position of the quantum entanglement peak is very close to but not coincident with the parameter value at which a phase transition occurs is not uncommon in 1D physical systems \cite{Schneider,Osterloh}. This is typically a finite size effect (i.e for systems that are not in the thermodynamic limit). However, for other systems, the entanglement peak is coincident with the parameter values at which a phase transition occurs: this has been shown in the Dicke model \cite{vidal:2006} and Lipkin model \cite{barthel:2006} in particular. See also the model in \cite{Lambert}.


We also calculate the purity of the quantum steady state (the trace of the square of the reduced density matrix at its steady state, $\tr{\te{\rh_0}^2}$). We compute the purity for each quantum steady state in a linear sweep of the parametric pumping $\ka$ whilst holding other parameters constant at the same values used to investigate the quantum steady state phase space in the previous section.

\begin{figure}[!htbp]\begin{center}
\includegraphics[scale=0.7]{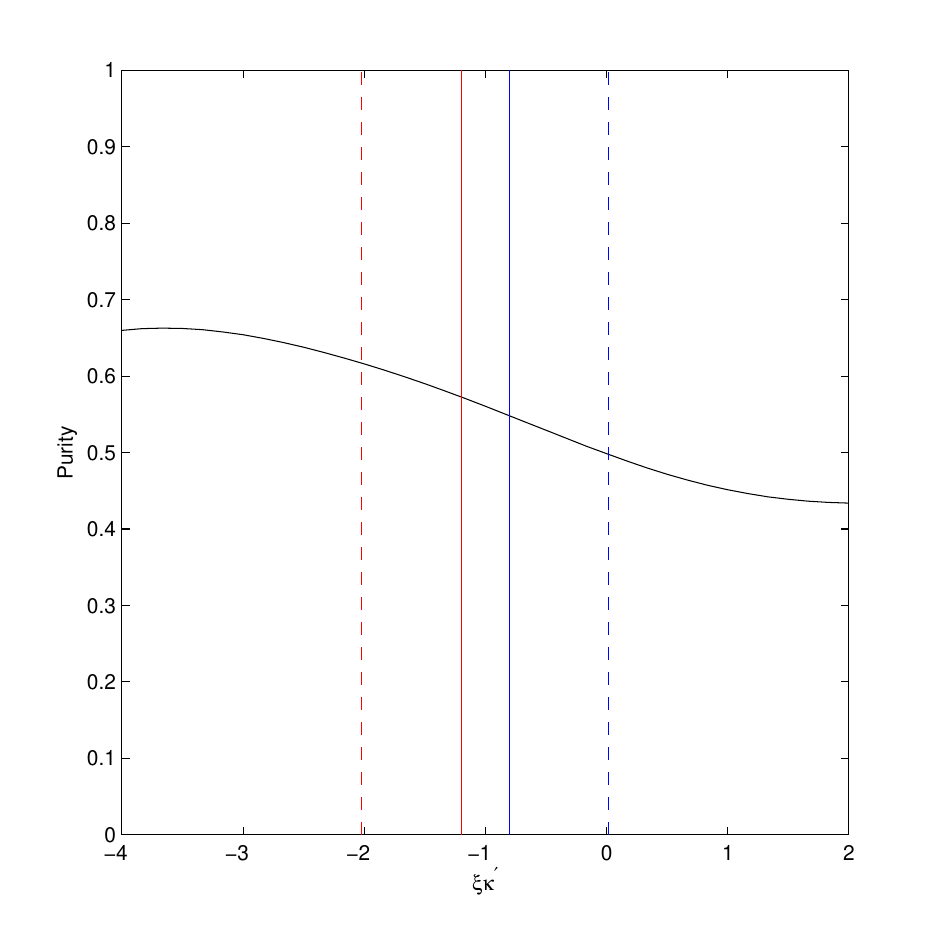}
\caption{(Color online) Purity of the steady state density matrix, for positive detuning ($\om>0$) through a linear sweep of increasing parametric pumping. The four vertical lines show the positions of the semi-classical bifurcations: the first vertical line (the dotted red line) is where the two unstable off-circle fixed points appear at the origin (the F-E boundary); the second vertical line (the solid red line) is where just created unstable fixed points annihilate the unit circle pair (the E-A boundary); the third vertical line (the solid blue line) is where the two pairs of fixed points re-appear (the A-G boundary); and the fourth vertical line (the dotted blue line) is where the off-circle pair vanishes at the origin (the C-D boundary). The purity is reduced as the separation of the Wigner function density increases. Backing out the original couplings necessary for the example parameter values plotted for, $\om=2,\ch=1,g=1.16,\mu=0.2,\ga=0.07$. The parametric pumping $\ka$ is linearly swept from $\ka=-8$ to $\ka=4$. In terms of the parametrisation used in the text, and in the phase diagram of Figure \ref{f:bif}, we have $\xi\ga\pr=0.2$ while $\xi\ka\pr$ is swept from $\xi\ka\pr=-4$ to $\xi\ka\pr=2$.}
\label{f:pp}
\end{center}\end{figure}

\begin{figure}[!htbp]\begin{center}
\includegraphics[scale=0.7]{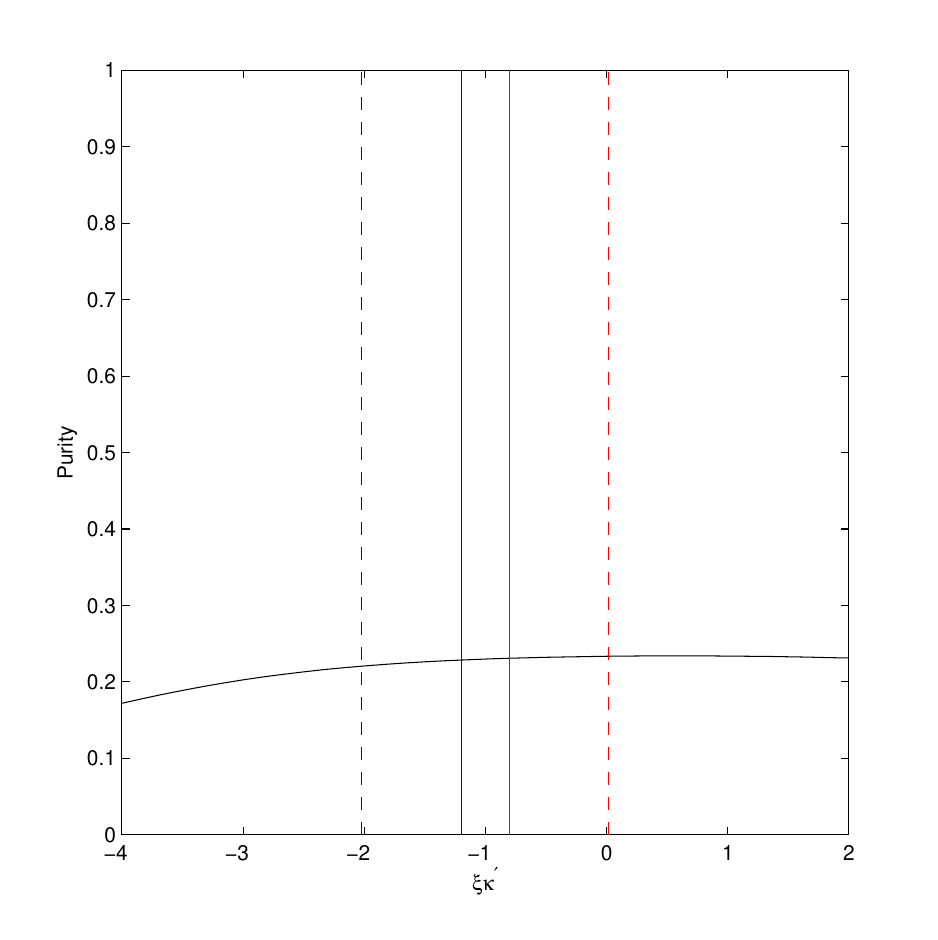}
\caption{(Color online) Purity of the steady state density matrix, for negative detuning ($\om<0$) through a linear sweep of increasing parametric pumping. The four vertical lines show the positions of the semi-classical bifurcations (from right to left): the first vertical line (the dotted red line) is where the two off-circle fixed points appear (the L-K boundary); the second vertical line (the solid red line) is where just created fixed points annihilate the unit circle pair (the K-J boundary); the third vertical line (the solid blue line) is where the two pairs of fixed points re-appear (the J-I boundary); and the fourth vertical line (the dotted blue line) is where the off-circle pair vanishes (the I-H boundary). The purity for negative detuning is remarkably flat. Backing out the original couplings necessary for the example parameter values plotted for, $\om=-2,\ch=1,g=1.16,\mu=0.2,\ga=0.07$. The parametric pumping $\ka$ is linearly swept from $\ka=-8$ to $\ka=4$. In terms of the parametrisation used in the text, and in the phase diagram of Figure \ref{f:bif}, we have $\xi\ga\pr=-0.2$ while $\xi\ka\pr$ is swept from $\xi\ka\pr=2$ to $\xi\ka\pr=-4$.}
\label{f:pn}
\end{center}\end{figure}

The purity for the positive and negative detuning cases are plotted in figures \ref{f:pp} and \ref{f:pn} respectively. We see that the purity is reduced when the separation of the Wigner function density from the origin increases. We interpret this as being consistent with the well-known notion\cite{CL,WM,zurek:1991} that as physical separation of two quantum objects increases, the decoherence is greater, and they will become more classical (less pure).

As mentioned, for ease of numerical simulation, we have picked parameters such that the steady state phase space density over our two sweeps of parameter values remains largely close to the origin in both the microwave and nano-mechanical spaces. That this means our sharp semi-classical bifurcations are more significantly blurred by quantum fluctuations is one reason we offer for the relatively small variation of both the entanglement and the purity over our chosen sweep.


\section{Conclusion.}

\label{s:c}

In this paper we detailed the semi-classical and quantum steady state structure of a particular dissipative nanomechanical system. In particular we observed that the semi-classical model contains a rich bifurcation structure, and that the remains of this structure are still visible in the full quantum mechanical steady state. The steady state quantum entanglement was found to be a maximum just to one side of the centre of the semi-classical bifurcation.

Specifically, the semi-classical fixed point structure contains not just one bifurcation, but a series of bifurcations, as a parameter is varied. The semi-classical fixed points' existence and locations were shown to be dependent on just two dimensionless parameter combinations, while their stability was shown to be dependent on these two together with an additional two dimensionless parameters. Semi-classically there were also periodic steady states for negative values of the detuning parameter.

The numerically calculated quantum steady states were shown to have clear signatures of these semi-classical steady state bifurcations. Specifically, the Wigner function representation of the quantum phase space was seen to have support on the semi-classical fixed points. In addition, where the semi-classical model had no stable fixed points, but instead had a periodic steady state, the Wigner function was seen to have support all around the limit cycle. This is consistent with the quantum phase space being completely phase-diffused around the limit cycle in the quantum steady state.

The ``Cassinian'' oscillator is thus an example of a correspondence principle between classical dynamics and quantum steady states. This principle is that investigation of the dynamics of the relevant semi-classical model gives significant predictive power for the steady state behaviour of the full quantum dissipative system.

Experimentally, variation of the parametric pumping $\ka$, whilst holding all other parameters constant and for a positively-detuning cavity drive $\om>0$, allows tuning through the semi-classical saddle-node bifurcations (from regions F to D in Figure \ref{f:bif}). Semi-classically, this means the fixed points switch from being aligned along the x-quadrature (for negative $\ka$) to being aligned the y-quadrature (for positive $\ka$). The calculated quantum steady-state phase space of section \ref{s:q} indicates that read-out of the microwave cavity steady state should see a splitting in the $x$ quadrature as the semi-classical bifurcation is crossed in accordance with Figure \ref{f:hqpc}. Such a splitting becomes sharper the closer in parameter space one is to the x-axis in the ``phase diagram'' of Figure \ref{f:bif} (where a smaller change in parametric pumping is required to transit regions E, A, G, B, and C). Experimentally, this means that the bifurcation is always achievable; however, the stronger the coupling $g$, the sharper the transition.

Currently achievable experimental values for such a nano-mechanical system are: a microwave cavity of frequency $\om_c=2\pi\times5\un{GHz}$ and a linewidth $\mu=2\pi\times490\un{kHz}$; a nano-mechanical resonator of mass $m=2\un{pg}$, a frequency under tension $\om_m=2\pi\times2.3\un{MHz}$, and a linewidth $\ga=2\pi\times20\un{Hz}$; and a microwave-mechanical coupling $G_0=2\pi\times1.16\un{kHz nm^{-1}}$ \cite{lehnert:2008}, which yields a linearised $g$ that can be increased by stronger driving up to a maximum circulating current of $1\un{\mu W}$ \cite{teufel:2008}. In fact the higher couplings $G_0=2\pi\times6.4\un{kHz nm^{-1}}$\cite{teufel:2008} and even $G_0=2\pi\times32\un{kHz nm^{-1}}$\cite{teufel:2009} have been achieved. We believe these values to allow the quantum signatures of the just mentioned semi-classical saddle-node bifurcations to be seen.

For negative detuning, and for some parameter regimes (some points in the meta-stable region A of Figure \ref{f:bif}), the nano-mechanics classically decays to periodic steady states, and the quantum system become completely phase-diffused, as in Figure \ref{f:hqn}. However, if read-out is to be via the microwave cavity, observation of these bifurcations will remain a challenge, since the phase-diffused rings are flattened as in Figure \ref{f:hqnc}. This is essentially because of some remnant of the correlation observed in \eqref{e:cncorrelation}.

In conclusion, even if the nano-mechanical experiment does not achieve a very large coupling, there are still semi-classical bifurcations which will be transited by varying the parametric pumping power. The overall transition from region F to region D of Figure \ref{f:bif} will leave a visible quantum signature which is the separation of steady state phase space density of the superconducting microwave cavity along one quadrature, as shown in Figure \ref{f:hqpc}. The detailed series of semi-classical bifurcations crossed in this transition will however be washed out. A stronger coupling will sharpen the quantum signatures of these semi-classical bifurcations.

\acknowledgments
This work has been supported by the Australian Research Council. Thanks to G. Vidal for useful discussions.


\end{document}